\documentclass{jfm}
\usepackage{hyperref} 
\usepackage{natbib}
\usepackage{upmath}
\usepackage[british]{babel}
\usepackage{amsmath,bm}
\usepackage{amssymb}
\usepackage{amsbsy}
\usepackage{amscd}
\usepackage{amstext}
\usepackage{tabularx}
\usepackage{float}
\usepackage{makeidx}
\usepackage{amsmath}
\usepackage{subfigure}
\usepackage{afterpage}
\usepackage[T1]{fontenc}
\usepackage[latin1]{inputenc}
\usepackage{multirow}
\usepackage{color}

\usepackage{graphicx} 
\usepackage{graphics,epsfig}
\usepackage{amsfonts}
\usepackage{psfrag}

\ifCUPmtlplainloaded \else
  \checkfont{eurm10}
  \iffontfound
    \IfFileExists{upmath.sty}
      {\typeout{^^JFound AMS Euler Roman fonts on the system,
                   using the 'upmath' package.^^J}%
       \usepackage{upmath}}
      {\typeout{^^JFound AMS Euler Roman fonts on the system, but you
                   dont seem to have the}%
       \typeout{'upmath' package installed. JFM.cls can take advantage
                 of these fonts,^^Jif you use 'upmath' package.^^J}%
      }
  \else
  \fi
\fi

\ifCUPmtlplainloaded \else
  \checkfont{msam10}
  \iffontfound
    \IfFileExists{amssymb.sty}
      {\typeout{^^JFound AMS Symbol fonts on the system, using the
                'amssymb' package.^^J}%
       \usepackage{amssymb}%
       \let\le=\leqslant  
         
      }{}
  \fi
\fi

\ifCUPmtlplainloaded \else
  \IfFileExists{amsbsy.sty}
    {\typeout{^^JFound the 'amsbsy' package on the system, using it.^^J}%
     \usepackage{amsbsy}}
    {\providecommand\boldsymbol[1]{\mbox{\boldmath $##1$}}}
\fi

\newsavebox{\astrutbox}
\sbox{\astrutbox}{\rule[-5pt]{0pt}{20pt}}

\def\u{{\hat{\mathbf{u}}}}
\def\tu{{\tilde{\mathbf{u}}}}

\def\pr{{\hat{p}}}
\def\tpr{{\tilde{p}}}

\def\q{{\hat{\mathbf{q}}}}

\def\U{{\mathbf{U}}}
\def\Ub{{\U_b}}

\def\Pb{{P_b}}

\newcommand\ad[2]{\left< #1, \ #2 \right>}

\usepackage{soul}

\title[RL-based control of confined cylinder wakes with stability analyses]
{Reinforcement-learning-based control of confined cylinder wakes with stability analyses}

\author[J. Li and M. Zhang]%
{
Jichao Li \and Mengqi Zhang\thanks{Email for correspondance: mpezmq@nus.edu.sg}
}

\affiliation{
 Department of Mechanical Engineering, National University of Singapore, 117575 Singapore 
  \\[\affilskip]
}

\date{\today}

\graphicspath{{figures/}}
\newlength\savewidth

\begin{document}
\maketitle

\begin{abstract}
This work studies the application of a reinforcement-learning-based (RL) flow control strategy to the flow past a cylinder confined between two walls in order to suppress vortex shedding. The control action is blowing and suction of two synthetic jets on the cylinder. The theme of this study is to investigate how to use and embed physical information of the flow in the RL-based control. First, global linear stability and sensitivity analyses based on the time-mean flow and the steady flow (which is a solution to the Navier-Stokes equations) are conducted in a range of blockage ratios and Reynolds numbers. It is found that the most sensitive region in the wake extends itself when either parameter increases in the parameter range we investigated here. Then, we utilise these physical results to help design RL-based control policies. We find that the controlled wake converges to the unstable steady base flow, where the vortex shedding can be successfully suppressed. A persistent oscillating control seems necessary to maintain this unstable state. The RL algorithm is able to outperform a gradient-based optimisation method (optimised in a certain period of time) in the long run. Furthermore, when the flow stability information is embedded in the reward function to penalise the instability, the controlled flow may become more stable. Finally, according to the sensitivity analyses, the control is most efficient when the probes are placed in the most sensitive region. The control can be successful even when few probes are properly placed in this manner.
\end{abstract}

\section{Introduction}

The flow past a cylinder constitutes one of the classical problems in fluid mechanics. 
It has intriguing flow phenomena that have attracted generations of researchers, hoping to unveil the essential fluid mechanics underneath~\citep{Williamson1996}. 
We consider here a specific type of the wake flow confined between two parallel walls~\citep{Schfer1996}. This flow is representative in many industrial and engineering scenarios such as a flow past dividers in polymer processing and turbulence promoters in the liquid-metal blankets of fusion reactors~\citep{Kanaris2011}. More importantly, we place the study of the confined wake flow in the context where we want to utilise a reinforcement-learning-based (RL) strategy to control the flow in order to evaluate the control performance of RL, possibly boosted by the flow physics knowledge we obtain from the stability analyses. Studies on machine-learning-based algorithms in fluid mechanics are burgeoning in recent years \citep{Duraisamy2019,Brunton2020}, but works on applying RL to flow control are still relatively few (see \cite{Rabault2020,Viquerat2021,Garnier2021} for a more complete literature review). There exist some unexplored topics in this field and some of them will be addressed in this work. In the following, we will first summarise the past works on confined and unconfined cylinder wake flows (especially their flow instability) and then review the recent development of machine learning in fluid mechanics (only relevant works will be discussed).

\subsection{Cylinder wake flows and their flow instability}
In the confined wake flow, the ratio of the cylinder diameter to the wall distance is termed as the blockage ratio~\citep{Chen1995,Sahin2004}.
\citet{Coutanceau1977} investigated this flow via experimental visualization techniques and identified the limits of the Reynolds number ($Re$) range in which the twin vortices exist and adhere stably to the cylinder.
\citet{Chen1995} showed that the formation of the steady vortex pair at the rear of the cylinder was not due to the bifurcation of the full dynamic system but instead it was probably associated with a bifurcation of a restricted kinematical problem.
They identified the Hopf bifurcation point by solving an eigenvalue problem resulted from linearisation,
and showed that the flow stability was lost through a symmetry-breaking Hopf bifurcation.
\citet{Anagnostopoulos1996} investigated the flows past a cylinder with three different blockage ratios at $Re=106$ (their description implied that they used the cylinder diameter as the reference length).
They found that the size of the standing vortices decreased with the blockage ratio before the Hopf bifurcation and 
the spacing of the vortices decreased with increasing blockage ratio when the wake became unsteady.
\citet{Sahin2004} systematically investigated two-dimensional flows past a confined circular cylinder with different blockage ratios.
The neutral stability curve was obtained via the linear global stability analysis.
They identified four regions in the parameter space of $Re$ (based on the cylinder diameter and maximum inlet fluid speed) and the blockage ratio, and each region corresponds to one type of flow motion:
steady symmetric flow, symmetric vortex shedding, steady asymmetric flow, and asymmetric vortex shedding.

As mentioned above, when the Reynolds number exceeds a critical value, the confined flows experience a Hopf bifurcation from a steady symmetric state towards a time-periodic non-symmetric state.
This is similar to a flow past an unconfined cylinder~\citep{Provansal1987,sreenivasan1987hopf}. Thus, it is necessary and instructive to review the works on the unconfined wake flow past a cylinder.
Continuous efforts have been made to understand the mechanism underneath the vortex shedding phenomenon, which is usually unwanted or even harmful.
It has been shown that the global instability~\citep{Noack1994} is responsible for the onset of the vortex shedding process~\citep{Jackson1987}.
\citet{Pier2002} showed that the two-dimensional time-periodic vortex shedding regime observed in the cylinder wake at moderate Reynolds numbers may be interpreted as a nonlinear global structure.
\citet{Barkley2006} studied the stability of a (time-)mean flow and showed that eigenfrequency of the mean flow tracked almost exactly the Strouhal number of the (nonlinear) vortex shedding.
Via a global weakly nonlinear analysis, \citet{Sipp2007} further confirmed for the cylinder flow that the mean flow was approximately marginally stable and showed that the linear dynamics of the mean flow yielded the frequency of the saturated Stuart-Landau limit cycle.
\citet{Leontini2010} conducted the linear global stability analysis on the mean flows and showed that the mean cylinder wake for $Re \le 600$ was marginally stable and the eigenfrequency of the leading global mode was close to the saturated vortex shedding frequency.

In addition to the linear stability analysis, sensitivity analysis has also been applied to the unconfined cylinder flows. Based on the insights provided by the stability analyses, \citet{Strykowski1990} managed to suppress the vortex shedding behind circular cylinders over a limited range of Reynolds numbers by a proper placement of a much smaller cylinder in the near wake of the main cylinder. Their results revealed that this part of the flow is important for flow control.
\citet{Hill1992} applied a sensitivity analysis based on the adjoint method~\citep{Jameson1988, Luchini2014} to the flow past a cylinder and computed the sensitivity of the least stable growth rate to the placement of a second smaller cylinder.
The sensitivity analysis reproduced the most sensitive regions that have been experimentally identified by~\citet{Strykowski1990}.
\citet{Giannetti2007} performed an analysis of the eigenvalue sensitivity to structural perturbations in the linearised governing equations and identified the `\textit{wavemaker}' region by overlapping the direct and adjoint perturbation modes,
which agreed well with the experimental data by~\citet{Strykowski1990}.
Using the linear stability theory and the adjoint method, \citet{Marquet2008} presented a general theoretical formalism to assess how base-flow modifications alter the flow stability.
\citet{Boujo2014} used the adjoint method to identify the regions that were the most sensitive to volume forcing and wall blowing/suction.
The control strategies designed based on the sensitivity analyses were proven to be effective via validations using the full nonlinear Navier-Stokes simulations.
Besides, sensitivity analyses have also been shown to be effective in the control of flow disturbances for the optimal transient growth~\citep{Corbett2001}, the noise amplification in a globally stable flat-plate boundary layer~\citep{Brandt2011}, and other flow control problems.
Nevertheless, since the adjoint-based optimal control law is obtained by minimizing the cost function such as time-averaged drag or flow fluctuation via the simulations of Navier-Stokes equations, 
it is still not computationally affordable to use this method in real-time active control of fluid flows. As an alternative, the RL control strategy will be studied in this work.

\subsection{Reinforcement learning as a flow control strategy}
Recently, reinforcement learning (RL), which has been used in some complex systems including automated driving and game playing, has been applied in the field of flow control.
Modelling the control as a Markov decision process, RL-based control agent is trained to take actions (to exert influences on the environment) to maximize the expected cumulative gains (reward) in a period.
RL can be treated as a black-box technique from the user side and establishes a control law from scratch.
\citet{Verma2018} used the RL-based control to find an efficient collective swimming strategy of fishes by harnessing vortices.
\citet{Rabault2019} applied reinforcement learning in active flow control for drag reduction in a confined wake flow at a moderate Reynolds number ($Re=100$ based on the averaged velocity and the cylinder diameter), and \citet{Rabault2019b} further presented a multi-environment approach to accelerate the training of the RL agent.
\citet{Tang2020} trained the RL agent to achieve a robust control of the drag reduction in the flow past the confined cylinder at multiple Reynolds numbers.
\citet{Xu2020} used RL-based control to stabilize the wake of the main cylinder by rotating two small cylinders located at two symmetrical positions downstream of the main cylinder.
\citet{paris2021robust} used a stochastic gated input layer in the RL agent to select an optimal subset from some initially placed probes.
\citet{Ren2021} performed a follow-up study of \cite{Rabault2019} and presented a successful application of the RL control in weakly turbulent conditions ($Re=1000$) with a drag reduction of $30\%$.
\citet{Beintema2020} applied RL in the suppression of Rayleigh-B{\'{e}}nard convection and discussed limitations in controlling unstable and chaotic dynamics.
Overall, most of the state-of-the-art works are focused on the validation of RL-based control in two-dimensional applications, which may not be persuasive enough for industrial applications in real-world flows with three-dimensional effects.
\citet{Fan2020} demonstrated, for the first time, the effectiveness of RL in experimental fluid mechanics by applying it in the drag reduction of circular cylinders in a turbulent flow.
The configuration chosen by~\citet{Fan2020} resulted in a simple control strategy (basically anti-clockwise rotations of control cylinders), so, regarding the use of RL for controlling turbulent flows, there would be much to investigate in more complex cases.

Recent developments of machine learning algorithms applied in fluid mechanics point to a very important line of research that we ought not to entirely rely on a brute-force strategy when designing and applying a machine learning algorithm in a flow problem, but to consider embedding some of the most fundamental physical or mathematical constraints or utilising some prior knowledge in the construction of the algorithm. 
This idea is drawing an increasing attention in a broad field of physics and engineering~\citep{vonRueden2021} as it will significantly reduce the searching space or guide the algorithm to advance in a more physically-relevant direction, helping to converge to the sought solutions more rapidly. 
For example, in the work of \cite{Ling2016}, a deep learning approach to RANS turbulence modelling that embedded Galilean invariance into the network using a higher-order multiplicative layer was presented.
This approach ensured that the predicted anisotropy tensor lies on an invariant tensor basis, and it was shown to have significantly more accurate predictions than a generic neural network that did not have any embedded invariance properties.
\cite{Raissi2019a} presented a physics-informed deep learning framework that synergistically combines mathematical models and training data, enabling scientific prediction and discovery from incomplete models and incomplete data.
\cite{Raissi2018} used the fluid mechanics governing equations as regularization mechanisms in the loss function of the deep learning network and demonstrated that this physics-informed deep learning algorithm is particularly effective for multi-physics problems such as vortex-induced vibrations of cylinders.
Similar applications can also be found in discovering turbulence models~\citep{Raissi2019c}, estimating hydraulic conductivity in Darcy flows~\citep{Tartakovsky2020}, and so on.
In the case of RL, \cite{Belus2019} found it useful to embed translational invariance into the architecture of the RL agent via the control of a one-dimensional depth-integrated falling liquid film.
\cite{Zeng2021} studied the Kuramoto-Sivashinsky equation using RL to minimize the dissipation rate and power cost in the chaotic system. Importantly, they trained the RL in a symmetry-reduced space \citep{Budanur2017}, showcasing the significance of considering embedding some physical constraint in the RL design.
We believe that there still remains a lot to do in the research of RL-based flow control that utilises prior knowledge of the flows in the construction of the algorithm. This is the theme of the current work. In particular, we will obtain the useful information provided by the flow stability and sensitivity analyses and use it in the RL-based control policies to suppress vortex shedding.
The flow instability mechanism, for example being absolutely or convectively unstable~\citep{Huerre1990}, affects the choice of control strategies.
Using the information of flow instability (coupled with linear control theory~\citep{Kim2007}) to modify the mean flow structure can be very efficient even by small-amplitude perturbations.
\citet{Delaunay1999} found that, in an unconfined flow past a cylinder, a slight suction destabilizes the wake in the subcritical $Re$ regime and a slight blowing stabilizes the flow in the supercritical regime.
As shown by~\citet{Sahin2004}, flows at different $Re$ and blockage ratios have such different characteristics that the optimal control strategy for them may differ.
Thus, being aware of some fluid information is helpful to design the control strategy.
In fact, in this work, we have experienced that directly applying RL to some challenging flow control problems (e.g., in the range of parameters that are difficult to control) without analysing the flow mechanism may fail.
The RL-based flow control by~\citet{Rabault2019} has shown its limitation in the cases where the drag reduction performance of the control policy becomes unstable~\citep{Tang2020}.
Analysing the instability mechanism may help to shed light on how to improve the control performance of RL and hopefully obtain more effective RL-based control strategies.
To the best of our knowledge, there are currently no studies in the literature on reporting how stability or sensitivity analysis can be effectively utilised in RL-based flow control.

\subsection{The position of the current work}
The primary aim of the current work is to explore the application of RL in fluid mechanics by harnessing the stability and sensitivity analyses in the RL-based control of confined wake flows. 
This parallels the efforts of embedding/utilising the flow physics in the machine learning studies to better leverage the power of the latter to obtain more physically-relevant results, exemplified in the important works such as \cite{Ling2016,Raissi2019a} among others, as reviewed above. 

The flow physics of the confined wake flow will first be investigated. We will reproduce some reported results on the global stability analysis of the confined wake flow as validation of our computations and will also apply the sensitivity analysis (which has not been applied to the confined wake flow) to describe the important flow structures/patterns and discuss their dynamics. These results will serve as a guidance for designing efficient control strategies in RL, which is the core theme of our study. In this part, a vanilla RL-based control method will first be applied to suppress the vortex shedding at different blockage ratios and Reynolds numbers.
The pros and cons of RL-based control in different regimes are analysed.
As a comparison, with the flow sensitivity used as a priori knowledge (for guiding the probe placement) and the stability information (embedded in the reward design), we will show that the performance of the RL-based control can be improved.

The paper is structured as follows.
In Sec.~\ref{problemformulation}, we introduce the confined cylinder wake problem and the control facilities.
In Sec.~\ref{secmethod}, we introduce the methodologies used in this work.
The results on the flow stability and sensitivity analyses are reported in Sec.~\ref{secResults}.
In Sec.~\ref{secDiscussion}, we discuss different facets of the RL-based control and present how to utilise the results of stability and sensitivity analyses to improve the control performance (especially the placement of the probes).
Finally, in Sec.~\ref{secConclusion}, we conclude the paper with some discussions.

\section{Problem formulation}
\label{problemformulation}

\begin{figure}
  \centering
  \includegraphics[width=0.7\textwidth]{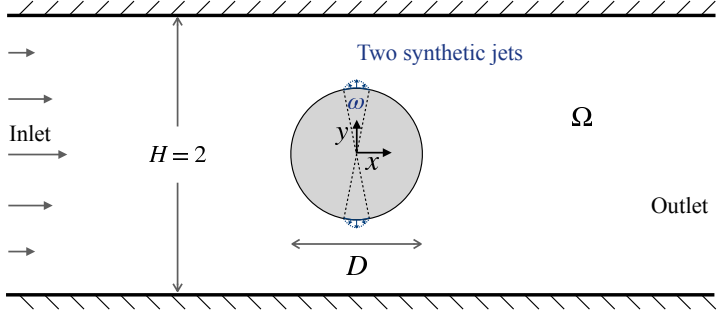}
  \caption{Cylinder symmetrically confined by two parallel no-slip walls. The channel height is $H=2.0$, and the cylinder diameter is $D$.
  The synthetic jet width is $w = \pi/18$.}
  \label{fg_geometry}
\end{figure}

We investigate the wake flow past a two-dimensional cylinder in a confined space~\citep{Chen1995,Sahin2004}, as shown in figure~\ref{fg_geometry}. We work with a Cartesian coordinate with $x$ and $u$ in the horizontal (or streamwise) direction and $y$ and $v$ in the vertical (or wall-normal) direction. The length is nondimensionalised by $D$ (which is the diameter of the cylinder) and the velocity by $U_{\text{max}}$ (which is the maximum velocity of the parabolic inflow, to be discussed below).
The cylinder is placed in the middle of the confined channel, and the coordinates of the cylinder center are $(0.0, 0.0)$.
We use the blockage ratio $\beta = D/H$ to quantify the degree of blockage.
The nondimensionalised inflow profile is given as:
\begin{equation}
\begin{aligned}
u(y) = 1-y^2, \ \ \ \ \ v(y) = 0.
\end{aligned}
\end{equation}
At the surface of the cylinder and on both sides of the channel, we apply the no-slip boundary condition.
On the right-hand side of the computational domain, we impose an outflow condition with 
$(p \boldsymbol{I} - \frac{1}{Re}  \nabla \boldsymbol{u} ) \cdot \boldsymbol{n}  = 0$, where $p$ is the pressure, $Re$ is the Reynolds number, $\boldsymbol{u}=(u,v)^T$ is the velocity, and $\boldsymbol{n}$ is the outward normal. 
The Reynolds number is defined as $Re= U_{\text{max}}D/ \nu$, where $\nu$ is the kinematic viscosity. Note that some of the previous works on the confined cylinder wake flow \citep{Chen1995,Sahin2004} have also used this definition of $Re$ or its variant. A comparison to their results will be made below.

When the Reynolds number exceeds a critical value (see Sec.~\ref{sec_cReNumber}), the confined cylinder wake flow starts to fluctuate and exhibit vortex shedding, which may excite a structural vibration.
With active suction or blowing of the synthetic jet flows \citep{Glezer2011,Rabault2019}, the stability of confined cylinder wake can be modified, and then the vortex shedding may be damped. As shown in figure~\ref{fg_geometry}, the two synthetic jets are placed at the top and bottom tips of the cylinder with a width of $w = \pi/18$.
For both jets, as in~\cite{Rabault2019}, a velocity boundary condition of a cosine-like profile is applied:
\begin{equation}
\begin{aligned}
  \boldsymbol{f}_{\theta, Q_\text{upper}}(x,y,t) &= (f_x, f_y)^T = Q_\text{upper}(t)\frac{\pi}{2\omega R^2} \cos{\left( \frac{\pi}{\omega} (\theta - 0.5\pi) \right)}(x,y)^T \\
  \boldsymbol{f}_{\theta, Q_\text{lower}}(x,y,t) &= (f_x, f_y)^T = Q_\text{lower}(t)\frac{\pi}{2\omega R^2} \cos{\left( \frac{\pi}{\omega} (\theta - 1.5\pi) \right)}(x,y)^T,
\end{aligned}
\end{equation}
where $\theta$ is the radian angular coordinate of an arbitrary point $(x,y)$ on the jets' surface; $f_x$ and $f_y$ are the velocity components along the $x$ and $y$ directions, respectively.
The flow rates of the upper and lower jets are controlled by changing the scaler values of $Q_\text{upper}$ and $Q_\text{lower}$, respectively.
The condition $Q_\text{upper} + Q_\text{lower} = 0$ is enforced to ensure that there is no additional mass added to the flow. An effective active control law of the synthetic jet flow rates is vital to the suppression performance, and we will use RL to learn such a control policy.

\section{Methodologies}
\label{secmethod}

\subsection{Direct numerical simulation}

Flow simulations are performed by solving the 2D incompressible Navier-Stokes equations in the computational domain $\Omega$:
\begin{equation}
  \begin{aligned}
    \frac{\partial \boldsymbol{u}}{\partial t} + \boldsymbol{u}\cdot \nabla \boldsymbol{u} = -\nabla p + \frac{1}{Re} \nabla^2 \boldsymbol{u}, \ \ \ \ \ \  \ \nabla \cdot \boldsymbol{u} = 0.
  \end{aligned}
\end{equation}
The open-source Nek5000 code developed by~\cite{nek5000} is used.
The spatial discretization in Nek5000 is based on the spectral element method (SEM). 
In each spectral element, the velocity space is represented by $N\text{th}$-order Legendre polynomial interpolants based on tensor-product arrays of Gauss-Lobatto-Legendre (GLL) quadrature points. 
The SEM has been shown to have little numerical dispersion and dissipation, which is important in the stability analysis.
We use the two-step backward differentiation formula for time integration in the unsteady flow simulation with a time-step of $5 \times 10^{-3}$ unit time. 
Based on the mesh convergence study (Sec.~\ref{sec_meshstudy}), we choose a mesh with 273 elements of order $N = 7$, which leads to 17472 grid points.
The flow field is advanced from a certain initial flow field; however, due to the convective effect in the flow, the initial conditions are not important in the following analysis and we will analyse and control the period of vortex shedding.
The lift force $F_L$ and drag force $F_D$ are computed by integrating forces on the cylinder surface, and the lift coefficient and drag coefficient are defined as $C_l=\frac{F_L}{{0.5DU_\text{max}^2}}$ and $C_d=\frac{F_D}{{0.5DU_\text{max}^2}}$, respectively.

\subsection{Linear stability analysis}

We will conduct the linear stability analysis to study the flow stability/instability of the confined wake flow. To linearise the incompressible Navier-Stokes equations, the total flow states ($\boldsymbol{u}$, $p$) are decomposed as a sum of steady base states ($\Ub$, $\Pb$) and infinitesimal perturbations ($\tu$, $\tpr$).
Based on the specific problems to be analysed below, $\Ub$ can be chosen as the mean flow or the steady-state solution to the nonlinear Navier-Stokes equations, which are called respectively the mean flow and the base flow in this work.
The mean flow can be easily obtained by time-averaging the DNS results. For the base flow, however, when the $Re$ number is greater than the critical $Re_c$, the cylinder wake flow experiences a Hopf bifurcation~\citep{Sahin2004} and evolves to be a time-periodic non-symmetric state.
Thus, the steady-state solution cannot be obtained by a time-marching method.
We use the selective frequency damping (SFD) method developed by~\cite{Akervik2006} to damp the unsteady temporal oscillations via a low-pass filter.

In the linear stability analysis, the perturbations are assumed to be in the form of normal modes $(\tu (x,y,t), \tpr (x,y,t))^T = (\u (x,y),\pr (x,y))^T\exp(\sigma t)$ with $\sigma= \lambda + i \omega$, where the real part $\lambda$ and the imaginary part $\omega$ are the growth rate and frequency of the mode, respectively, and $(\u (x,y),\pr (x,y))^T$ are called shape functions of the variables. After substituting the normal-mode ansatz into the Navier-Stokes equations and linearising them around the base state ($\Ub$, $\Pb$), we obtain
\begin{equation}
  \begin{aligned}
    \sigma \u + \nabla \u \cdot \Ub + \nabla \Ub \cdot \u &= - \nabla \pr + \frac{1}{Re} \nabla^2 \u, \  \ \ \ \   \nabla \cdot \u &= 0.
  \end{aligned}
\label{eq_direct}
\end{equation}
The boundaries for the linear direct problem are the same as those in the nonlinear Navier-Stokes equations, except that the inlet boundary condition for the velocity is a Dirichlet type with $\u = \boldsymbol{0}$.

The equations in Eq.~\ref{eq_direct} lead to an eigenvalue problem and the solutions constitute linear global modes of the problem.
For clarity, we use $\q$ to represent $(\u (x,y),\pr (x,y))^T$.
Then, the stability analysis can be investigated by solving the following eigenvalue problem
\begin{equation}
  \boldsymbol{A} \q = \sigma \q,
  \label{eq_linear} \ \ \ \  \ \text{with} \ \ \ \
    \boldsymbol{A} = \begin{pmatrix}
   - \Ub \cdot \nabla - \nabla \Ub \cdot + Re^{-1} \nabla^2  & - \nabla \\
    \nabla \cdot & \boldsymbol{0} \\
   \end{pmatrix}.
\end{equation}
As one can see, the Jacobian matrix $\boldsymbol{A}$ depends on the base state ($\Ub$,$\Pb$).
This global eigenvalue problem can be solved by an iterative approach, and the most popular one is the Arnoldi algorithm~\citep{Arnoldi1951,Saad1980}. It is a time-stepping-based Jacobian-free method (meaning that one does not need to explicitly construct $\boldsymbol{A}$) and has been widely used in the global stability analyses of complex flow problems, e.g., \cite{Eriksson1985,Tezuka2006,Barkley2008} among many others, see also the review paper by \cite{Theofilis2011}. The major step is the generation of a Krylov subspace $\mathcal{K}_m$ by marching the linearised Navier-Stokes equations from a certain initial snapshot $\q_0$ at successive equidistant instants of time ($\Delta t$). An orthogonal basis is then generated by the Gram-Schmidt procedure, transforming the large-scale eigenvalue problem to a smaller one of Hessenberg form that can be solved easily. 

\subsection{Sensitivity analysis}

The sensitivity analysis based on the adjoint method will also be performed in the current work. It is an important tool which has been extensively applied in flow control and shape optimisation.
As in~\cite{Giannetti2007,Marquet2008}, the adjoint equation of the linearised Navier-Stokes equations reads
\begin{equation}
  \begin{aligned}
  \sigma^* \u^+ - \nabla \u^+ \cdot \Ub + (\nabla \Ub)^T \cdot \u^+ &= - \nabla \pr^+ + \frac{1}{Re} \nabla^2 \u^+, \ \ \ \ \   \nabla \cdot \u^+ &= 0,
\end{aligned}
\label{eq_adjoint}
\end{equation}
where $\u^+$ and $\pr^+$ are the adjoint vectors to $\u$ and $\pr$, respectively.
In principle, the boundary conditions for the adjoint equation are (following \cite{Peplinski2014})
\begin{subeqnarray}\label{adjoint_bc}
  \u^+ = \boldsymbol{0} &&\quad \text{at the inlet and walls}, \\
  \pr^+ \boldsymbol{n} - Re^{-1} (\nabla \u^+) \cdot \boldsymbol{n} = (\Ub \cdot \boldsymbol{n})\u^+ && \quad \text{at the outlet}.
\end{subeqnarray}
The boundary conditions \ref{adjoint_bc}b is not supported in the current SEM flow solver \citep{Peplinski2014}. Instead \cite{Giannetti2007} explained that because of the particular structure of the base flow, the adjoint mode decays rapidly away from the cylinder; therefore, $\u^+ \rightarrow \boldsymbol{0}$ can be considered when the outlet is far enough from the cylinder. This method is adopted here because a far enough outlet has been used in the current work (see the geometry in figure \ref{fg_mesh} in the following) and we have checked that in our simulations, the amplitude of $\u^+$ at the outlet is almost zero.

Similar to the direct problem, the adjoint can be solved by the Arnoldi method.
For certain flows, the structural sensitivity analysis can be used to locate the origin of the instability perturbations, called the \textit{wavemaker} region, which can help to understand the instability mechanism~\citep{Pier2001}.
As shown in \cite{Giannetti2007}, the \textit{wavemaker} region $\boldsymbol{\eta}$ can be identified by overlapping the direct eigenvector $\u$ and adjoint eigenvector $\u^+$
\begin{equation} 
\label{eq_wavemaker}
  \boldsymbol{\eta} = \frac{|\u||\u^+|}{\ad{\u}{\u^+}}.
\end{equation}

It is noted that for the confined cylinder wake flow, we cannot find results on its sensitivity analysis in the literature. Thus, the results reported below on this analysis will be interesting by themselves (especially, the variation of the \textit{wavemaker} region when $Re$ or $\beta$ changes). On the other hand, the linear stability analyses of confined cylinder wake flows have been documented \citep{Chen1995,Sahin2004} and we will compare our results (based on the SFD base flow) with them for the validation purpose and will further perform the stability analysis based on the mean flow. 

\subsection{Reinforcement Learning}

\begin{figure}
  \centerline{\includegraphics[trim=0cm 0.1cm 0cm 0cm, width=0.9\textwidth]{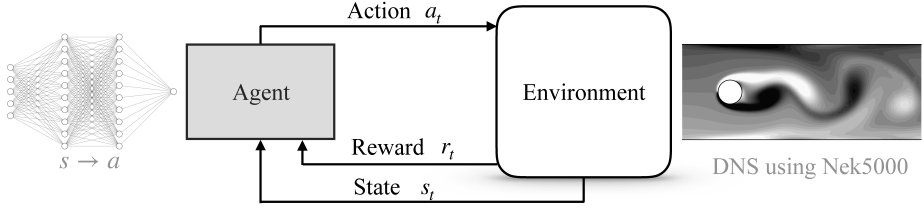}}
  \caption{The reinforcement learning framework in flow control. Agent: neural network; Environment: DNS using Nek5000; Action: adjustment of the synthetic jet flow rates; Reward: reduction of the shedding energy; States: spatial velocities.}
  \label{fg_RL}
\end{figure}

Reinforcement learning trains the control agent from scratch by interacting with the environment and maximizing the expected cumulative reward.
As shown in figure~\ref{fg_RL}, reinforcement learning is composed of three fundamental components, i.e., the agent, the environment, and the reward function.
The agent usually contains a neural network such as a multilayer perceptron or convolutional neural network that is used to determine the control action $a_t$ based on the current state of the environment $s_t$;
the action is then applied to the environment;
and the reward $r_t$ (evaluating the quality of control actions) is calculated and recorded for network updates.

As a policy-gradient method, the RL agent network (with parameters $\theta$, i.e., weights and biases) is trained to find the optimal policy $\pi_{\theta}\left(a_t | s_t\right)$, which is the distribution probability of action $a_t$ (with respect to the states $s_t$) to maximize the expected cumulative reward:
$R_t = \sum_{k>t} \gamma^{(k-t)}r_t$, where $\gamma \in (0,1]$ is a discount factor.
The current RL agent uses the proximal policy optimisation (PPO) method developed by~\cite{schulman2017proximal} to update the parameters $\theta$.
PPO is an episode-based actor-critic algorithm.
In addition to the network approximating the policy $\pi$ for action distributions (called `actor'), PPO involves a critic network $V$ to predict the discounted reward with respect to the states $s_t$, which is further used to update actor network.
When training the critic network, $\hat{A_t} = R_t - V(s_t)$ is defined to measure the discrepancy between the predicted and actual discounted rewards, and the loss function to be minimized can be defined as $L_\text{critic} =\hat{\mathbb{E}_t} (\hat{A_t}^2)$, where $\hat{\mathbb{E}_t}$ is the empirical expectation over time.
A clipped surrogate objective function is maximized to update the actor network:
$L^\text{Clip} (\theta) = \hat{\mathbb{E}_t} \left[ \min\left(p_t(\theta)\hat{A_t}, \text{clip}\left( p_t(\theta), 1-\epsilon, 1+ \epsilon\right)\hat{A_t} \right) \right] $, where $p_t(\theta) = \pi_{\theta}(a_t | s_t) / \pi_{{\theta}_\text{old}}(a_t | s_t)$.
The clip term removes the incentive for moving $p_t$ outside of the interval $[1-\epsilon,1+\epsilon]$ ($\epsilon$ is 0.2 as recommended) and thus prevents excessively large policy update.
More technical details on PPO can be found in~\cite{schulman2017proximal}.
The adam optimizer is used to update the networks and the learning rate is fixed as 0.001.

In this work, for the suppression of the vortex shedding in the confined wake flow past a cylinder, the RL environment is simulated by the direct numerical simulation of the wake flow using Nek5000.
Referring to the open-source RL-based cylinder flow control repository developed by~\cite{Rabault2019} based on the Tensorforce library~\citep{tensorforce}, we present an open-source Python code to interface the Nek5000 simulation environment with the RL agent, which is available as a GitHub repository\footnote{\url{https://github.com/npuljc/RL_control_Nek5000}}.
Since the flow rates of two synthetic jets are confined according to $Q_\text{lower} + Q_\text{upper}=0$,
the RL action is realised by manipulating the flow rate of the upper synthetic jet, and the lower synthetic jet has the same flow rate but opposite direction.
Vortex shedding frequencies of the cases studied in this work are about $0.3 \sim 0.4$ (see Sec.~\ref{sec_vortex}), and we choose a duration of $\Delta t = 0.2$ between two control actions (corresponding to $6\% \sim 8\%$ vortex shedding period) to leave a large degree of control freedom. 
We define a training episode composed of 16 time units, which is corresponding to $4.8 \sim 6.4$ vortex shedding periods, and thus, 80 actions will be taken in each episode.
To avoid abrupt changes, we adopt the same strategy used by~\citet{Rabault2019} to gradually update the jet flow rate after each time step in DNS; that is, $u_\text{jet}^{(t+1)} = u_\text{jet}^{(t)} + 0.1\times (u_\text{action} - u_\text{jet}^{(t)})$.
Following~\citet{Rabault2019}, we define the policy network as a multilayer perceptron with two hidden layers $(512\times512)$.
The RL agent is updated every 20 episodes in the training process.
Probe sensors that monitor velocity components in both directions are placed in the flow field to provide environment states for RL.

\citet{Rabault2019} discussed that the number of probes has a direct influence on the control performance in RL. 
We extend this investigation and further determine that the probes are better placed in the regions that are important in the sensitivity analysis. This heuristic approach may also be helpful to be combined with the optimal searching method proposed by \cite{paris2021robust}.
Besides, the definition of the reward function is important in RL-based control.
\cite{Rabault2019,Rabault2019b} used a drag-based reward function in order to control (reduce) the drag force on the cylinder.
In the current work, in order to damp the vortex shedding, we use a reward function defined based on the kinetic energy of vortex shedding, and the details are discussed in Sec.~\ref{sec_reward} and Sec.~\ref{sec_rewardSFD}. 

\section{Results: stability and sensitivity analyses of confined wake flows}
\label{secResults}

\subsection{Mesh convergence study}
\label{sec_meshstudy}

\begin{figure}
\centerline{\includegraphics[width=\textwidth]{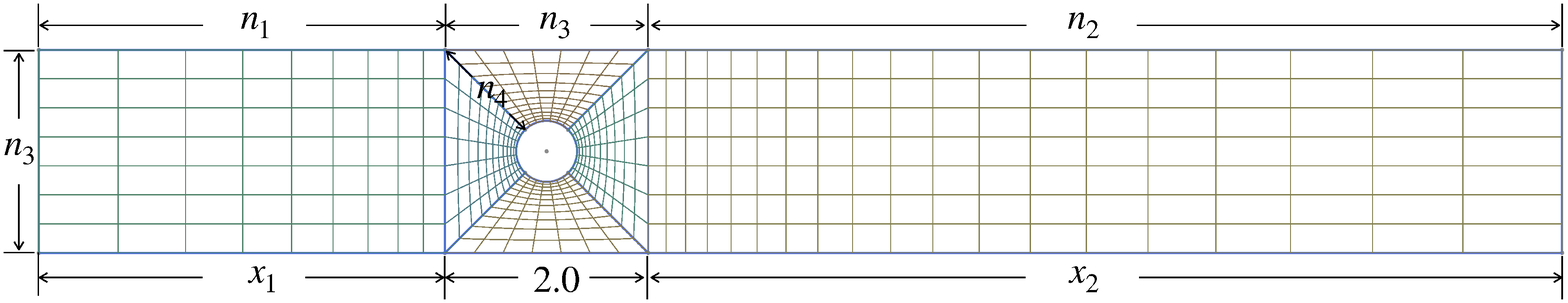}}
\caption{The computational mesh is composed of two rectangle domains and a square domain. }
\label{fg_mesh}
\end{figure}

As shown in figure~\ref{fg_mesh}, the computational domain is defined by three parts for the ease of mesh generation.
$x_1$ and $x_2$ are the lengths of the two rectangle subdomains in the streamwise direction.
The two rectangle subdomains are discretised using $n_1 \times n_3$ and $n_2 \times n_3$ element nodes, respectively.
The middle square domain with a side length of 2.0 is discretised using an ``O''-type mesh with $4 \times n_3 \times n_4$ element nodes (4 here denotes the four compartments delimited by the blue lines in the figure; $n_3=8$ and $n_4=6$).
In order to obtain a reasonable computational mesh, we study the influences of both the computational domain size and the mesh resolution on the numerical results.
As shown in Table~\ref{tb_grid}, five computational sizes are investigated for $Re=200, \beta=0.5$ ($Re=200$ is in the upper limit of the Reynolds numbers we investigate in this work), and $E$ is the total number of elements.
For each computational size, we compare numerical results by using meshes of two resolution levels (L1 and L2).
The L1 mesh is finer, which is generated by doubling the element numbers used in the L2 mesh.
The time-averaged $C_d$ evaluated by different meshes are shown in Table~\ref{tb_grid}, and the reference value evaluated by~\citet{Sahin2004} is 2.4245.
The choice of five computational domains does not significantly influence the numerical results and all are close to the reference.
This means that using the smallest computational domain (D1) is sufficient.
Comparing with the finer L1 mesh, the L2 mesh merely introduces an error of less than $0.003\%$.
In the following, we will use D5-L1 as the configuration for all the stability analyses below and in the case of RL control, in order to reduce the computational burden, we use D1-L2.

\begin{table*}
  \centering
  \caption{Resolution parameters of L2 meshes in five computational domains and the time-averaged $C_d$ on the confined cylinder with $\beta=0.5$ and $Re=200$ evaluated by different meshes. $E$ is the total number of elements in the spectral element method. }
  \begin{tabular}{c*{10}{r}}
  \hline
  Mesh  &    $x_1$    &  $x_2$    &     $n_1$    &  $n_2$  &   $E$   & $\overline{C_d}$ by L2  & $\overline{C_d}$ by L1  \\
  \hline
  D1      &     3.0       &   10.0      &    6      &   15      &  273   &  2.422105  & 2.422176  \\
  \hline
  D2      &     3.0       &   20.0      &    6      &   20      &  308   & 2.422132 & 2.422176  \\
  \hline
  D3      &     10.0       &  10.0      &    15      &  15      &  336  &  2.422096  & 2.422168 \\ 
  \hline
  D4      &     10.0       &  20.0      &    15      &  20      & 371   & 2.422139 & 2.422168 \\ 
  \hline
  D5      &     20.0       &  20.0      &    20      &  20      & 406   & 2.422140 & 2.422168 \\ 
  \hline
  \end{tabular}
  \label{tb_grid}
\end{table*}

\subsection{Critical Reynolds numbers}
\label{sec_cReNumber}

The dynamics of the confined cylinder wake flow is governed by the Reynolds number and the blockage ratio.
The vortex shedding in the wake flow occurs through a symmetry-breaking Hopf bifurcation beyond the critical Reynolds number ($Re_c$).
In the confined flow past a cylinder, $Re_c$ varies with the blockage ratio.
In order to determine the vortex shedding region in the $Re-\beta$ plane, we solve the eigenvalue problem in the linear stability analysis based on the SFD base flow.

\begin{figure}
\centerline{\includegraphics[width=\textwidth]{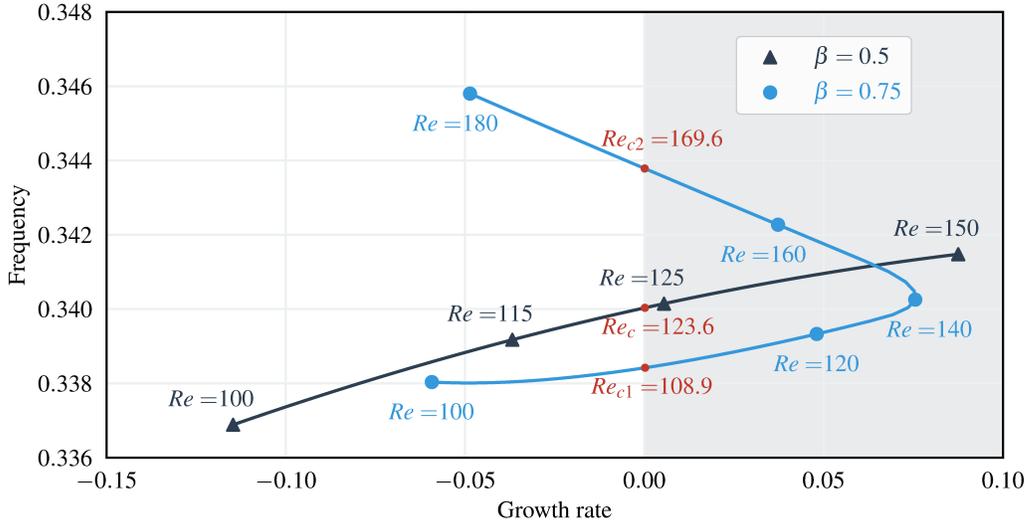}}
\caption{Eigenvalues of the leading eigenmode at different Reynolds numbers for two blockage ratio $\beta=0.5,0.75$. The curves are interpolated using data of the scatter points. The gray shade indicates flow instability. Note that that the data points are our computational results and the lines are fitted to guide the eyes.}
\label{fg_eigRe}
\end{figure}
  
\begin{figure}
\centering
\subfigure[$\beta=0.25$]{
    \includegraphics[width=0.32\textwidth]{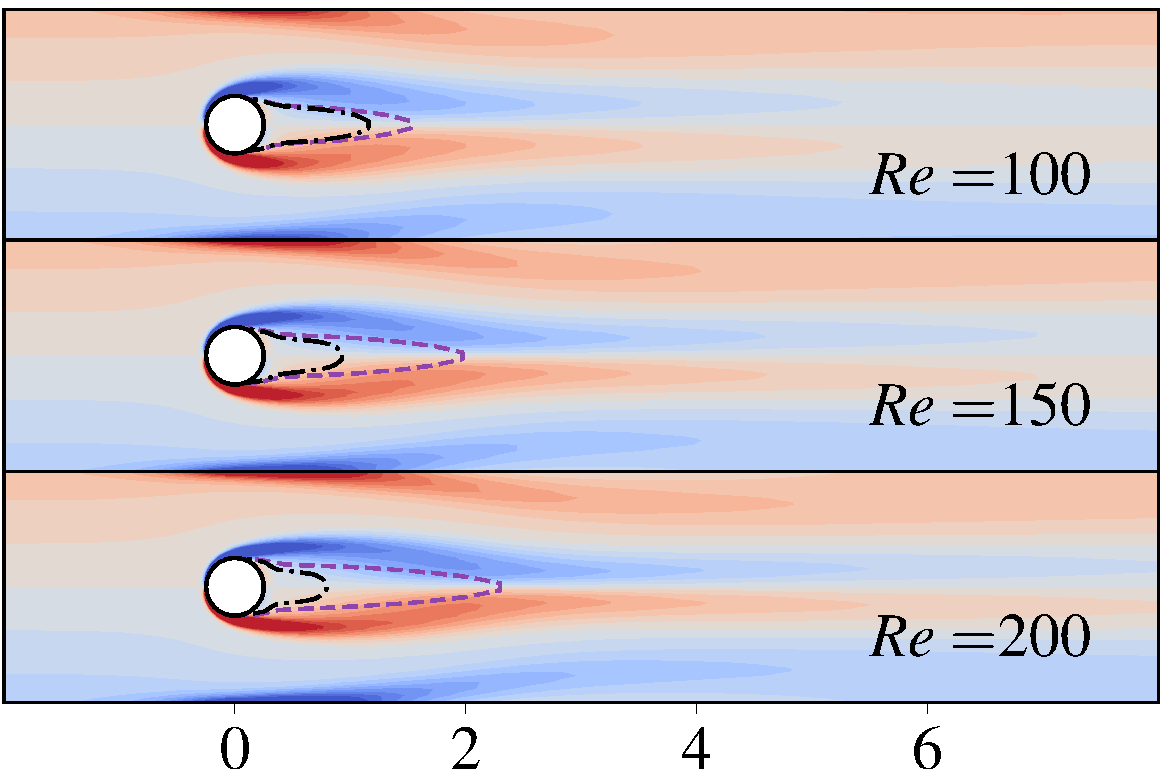}}
\subfigure[$\beta=0.5$]{
    \includegraphics[width=0.32\textwidth]{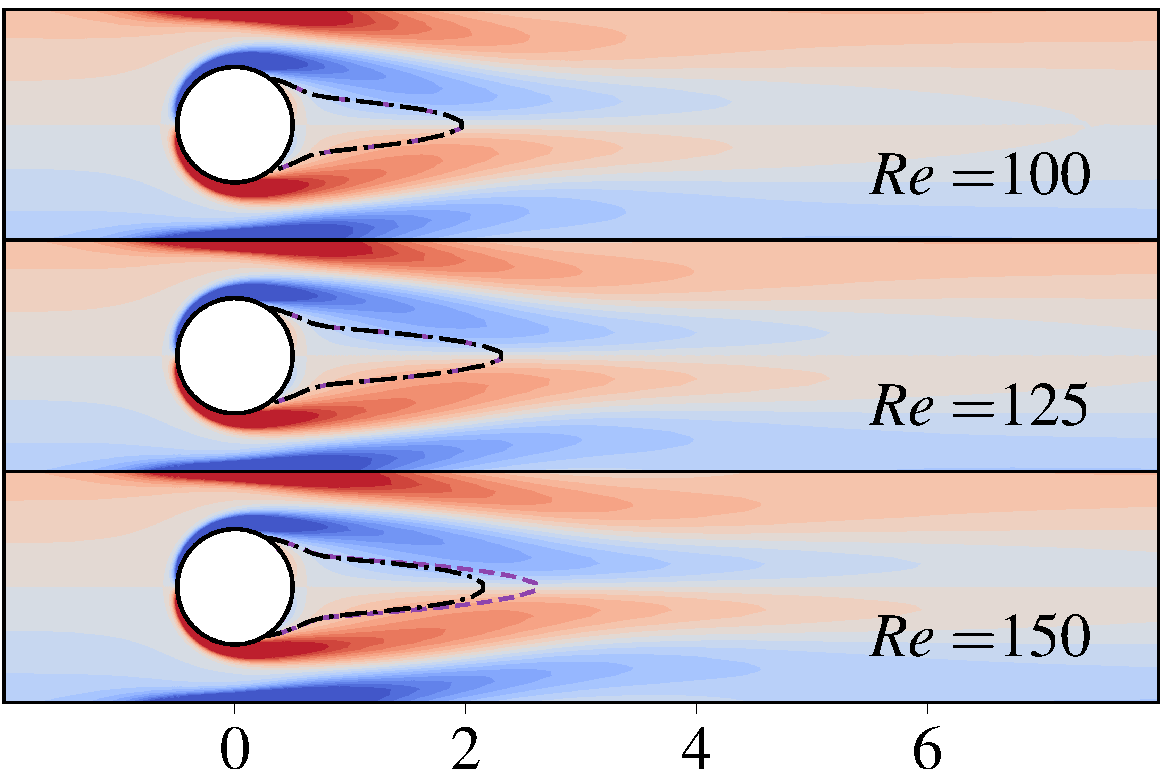}}
\subfigure[$\beta=0.75$]{
    \includegraphics[width=0.32\textwidth]{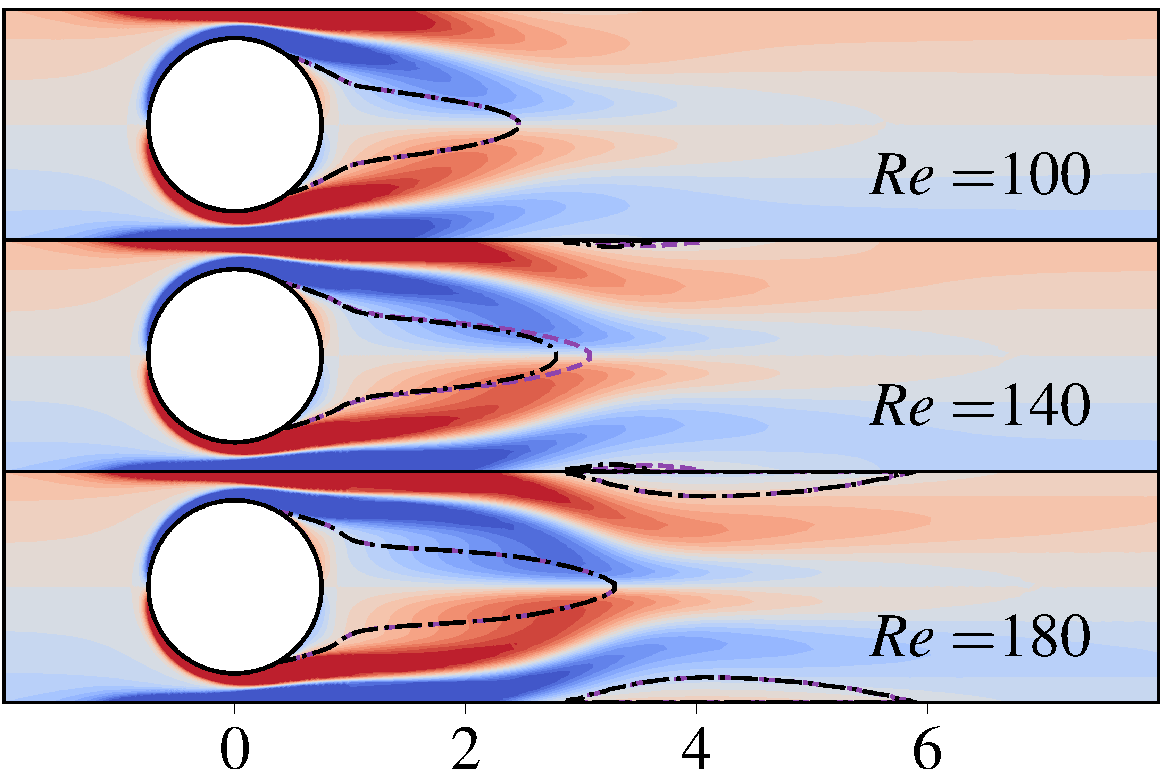}}
    \caption{Vorticity of SFD base flows of different confined cylinder wake flows. The purple dashed lines and black dash-dot lines are the recirculation zones boundaries of the SFD base flows and time-mean flows, respectively.}
    \label{fg_baseflows}
\end{figure}

As shown in figure~\ref{fg_eigRe}, two values of $\beta$ are studied. For the confined flow with $\beta=0.5$, the growth rate of the leading eigenmode increases monotonically with a rise of $Re$ from 100 to 150.
When the Reynolds number is greater than $Re_c = 123.6$, the base flow becomes unstable and then the wake starts meandering.
However, for the confined flow with the blockage ratio $\beta=0.75$, two critical $Re_c$, $Re_{c1} = 108.9$ and $Re_{c2} = 169.6$, are identified.
With the increase of $Re$ from 100, the flow stability is lost via the first Hopf bifurcation point at $Re_{c1}$ and the wake vortex starts shedding subsequently.
If the $Re$ further increases, passing the other bifurcation point at $Re_{c2}$, the flow gets stabilised.
As shown in figure~\ref{fg_baseflows}, the confined flows with $\beta=0.25$ and $\beta=0.5$ merely have one recirculation zone; whereas for the larger blockage ratio $\beta=0.75$, in addition to the recirculation bubble just downstream the cylinder, two additional recirculation bubbles develop close to the walls further downstream when $Re$ is large enough.
Similar to observations by~\citet{Sahin2004}, the recirculation bubbles on the confinement walls become larger with the increase of $Re$ and lead to the other Hopf bifurcation point at $Re_{c2}$.

\begin{figure}
\centerline{\includegraphics[width=\textwidth]{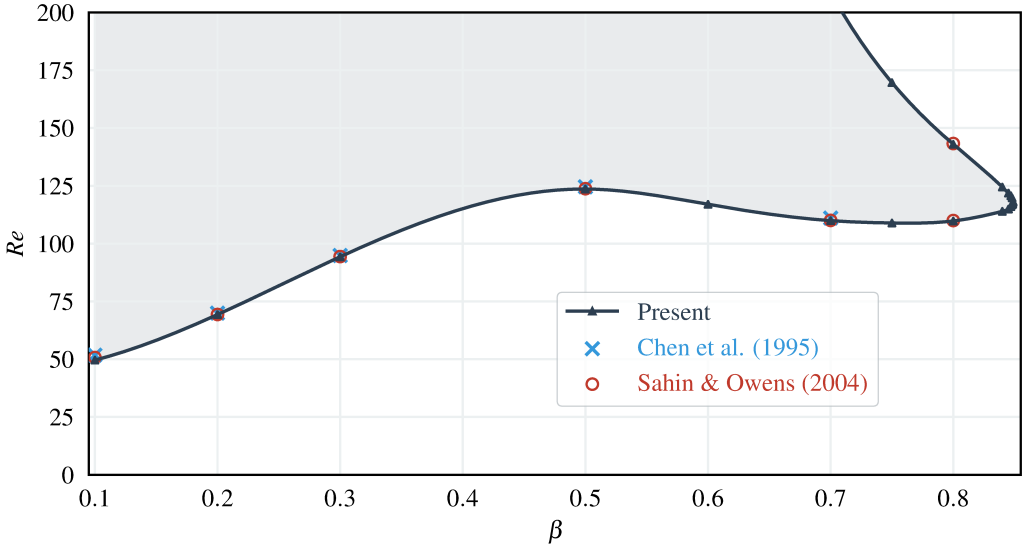}}
\caption{Vortex shedding region (gray) delimited by the neutral stability curve. Flows in the white region are stable.  A comparison is made to two previous publications as shown in the legend.}
\label{fg_Recs}
\end{figure}

By examining the critical Reynolds numbers with different blockage ratios, the vortex shedding region is shown in figure~\ref{fg_Recs} as a grey background in the $Re-\beta$ plane.
The region is determined by two branches of critical points, and we have used 13 and 8 data points in the lower branch and upper branch, respectively.
The favourable agreement with the results by \citet{Chen1995} and \citet{Sahin2004} demonstrates the good accuracy of the numerical methods and meshes used in this work.
The confined flow becomes more stable as $\beta$ increases up to 0.5.
Then, the flow gets destablilised as the block ratio increases.
When $\beta$ is larger than $\sim0.85$, the flow is again stabilised. 
For a relatively large $\beta$ ($\lessapprox 0.85$), another stability region can be identified when $Re$ is sufficiently large (when $Re\le200$).
It is noted that the flow phenomena are richer and more complex in the upper-right region of the $Re-\beta$ plane (see figure 5 of \cite{Sahin2004}) and they are not studied here. We will focus on the gray region where vortex shedding occurs and will use RL control to abate it.

\subsection{Vortex shedding phenomenon}
\label{sec_vortex}

In general, the oscillating Karman vortex street downstream the cylinder wake occurs due to the loss of instantaneous reflection symmetry through a Hopf bifurcation.
For the unconfined flow past a cylinder, it is suggested by \cite{Maurel1995,Noack2003} that the amplitude of the oscillating wake saturates when the time-averaged flow (mean flow) is marginally stable and 
the mechanism for nonlinear saturation of the oscillating wake flow is the mean flow correction/modification through the formation of Reynolds stresses.
The mean flow provides a good profile to predict the shedding frequency of the unconfined cylinder wake \citep{Yang1989,Pier2002,Barkley2006}.
In order to understand in detail the stability property of the confined cylinder wake, hereafter, we perform the global linear stability analysis of its mean flow. 

\begin{figure}
\centerline{\includegraphics[width=\textwidth]{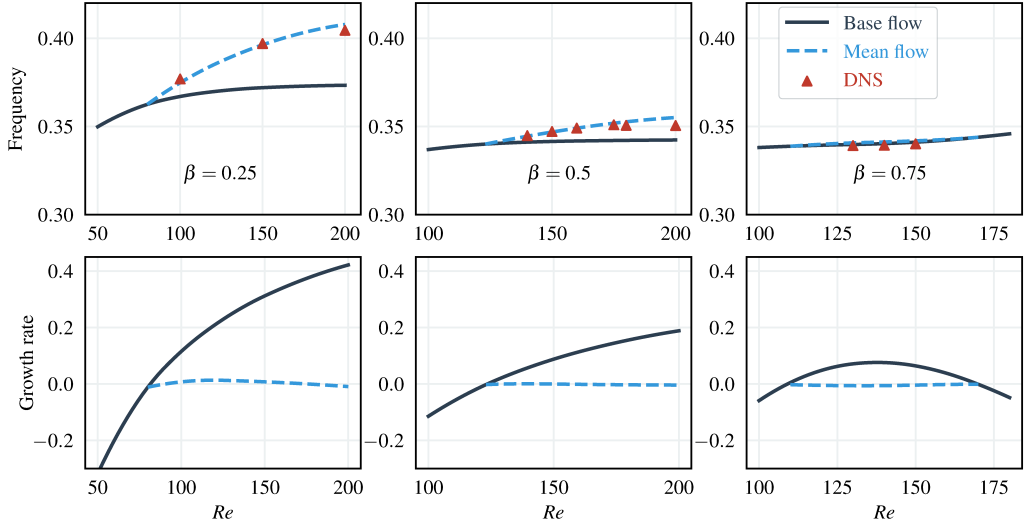}}
\caption{Frequencies and growth rates as a function of $Re$ in the global linear stability analysis of confined cylinder flows for three values of confinement ratio $\beta$.}
\label{fg_meanvsbase}
\end{figure}

Figure~\ref{fg_meanvsbase} shows the leading eigenvalues of the confined cylinder wake with $\beta=0.25$, $\beta=0.5$ and $\beta=0.75$.
Compared to the stability analysis based on the SFD base flow (black solid lines), the frequencies solved by the linear analysis based on the mean flow (blue dashed lines) agree better with the results of nonlinear DNS (red filled triangles),
and the relative discrepancies almost are within 1\%. 
The growth rates of the mean flows are close to zero, which implies that the confined mean flows are marginally stable. 
The conclusions hold for both the weak and strong confinement cases.
This implies that the confined cylinder wake flows approximately have the real-zero imaginary-frequency (RZIF) property~\citep{Turton2015}, which is similar to the unconfined cylinder wake. The RZIF property implies that the eigenfrequency of a nonlinearly-saturated oscillating flow can be well approximated by a linear analysis based on the time-mean flow \citep{Pier2002,Barkley2006,Sipp2007}. 
Besides, the linear stability analysis using the SFD base flow (black lines) generates apparently different results than those using the mean flow and of DNS (except the frequency in the case of $\beta=0.75$).

\begin{figure}
\centerline{\includegraphics[width=\textwidth]{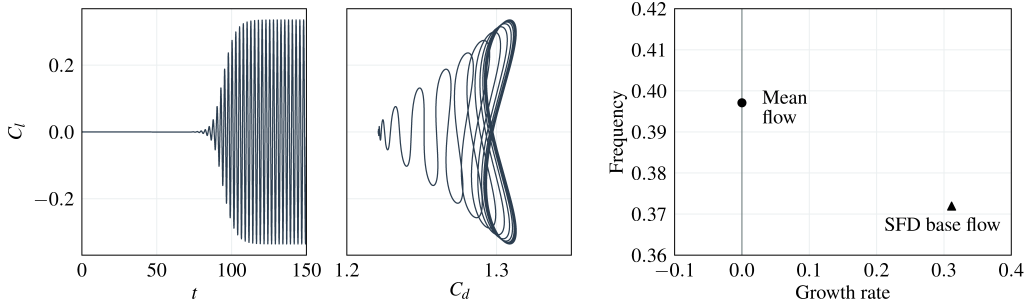}}
\caption{Relationship between the SFD base flow and the mean flow at $Re=150$ with $\beta = 0.25$. Left: $C_l$ as function of $t$ from the SFD base flow to the saturated flow with vortex shedding.
Middle: phase diagram of $C_l$ and $C_d$ from $t=0$ to $t=150$. 
Right: growth rates and frequencies of the SFD base flow and the mean flow solved by the linear stability analysis. 
}
\label{fg_base2mean}
\end{figure}

In order to reveal the relationship between the SFD base flow and the mean flow, we perturb and evolve the SFD base flow at $\beta=0.25$ at $Re=150$ to see how it develops to the saturated state (with vortex shedding). This analysis follows that for the unconfined wake flows in \cite{Barkley2006}.  
As shown in figure~\ref{fg_base2mean}, since the SFD base flow is unstable, the amplitude of the $C_l$ oscillation increases and eventually saturates at a periodic vortex shedding state.
The cone-like shape of drag evolution has been explained by~\cite{Loiseau2018} based on the sparse identification of nonlinear dynamics, SINDy \citep{Brunton2016}. Similarly to the unconfined flow past a cylinder, the evolution of the unstable SFD base flow in the confined wake is a nonlinear saturation of oscillations (as shown in the left and middle panels), and the selection of vortex shedding amplitude and frequency is based on the marginal stability of the mean flow.
In the right panel of figure~\ref{fg_base2mean}, we show the growth rates and frequencies of the SFD base flow and the saturated mean flow.

\subsection{Structural sensitivity}

\begin{figure}
\centerline{\includegraphics[width=\textwidth]{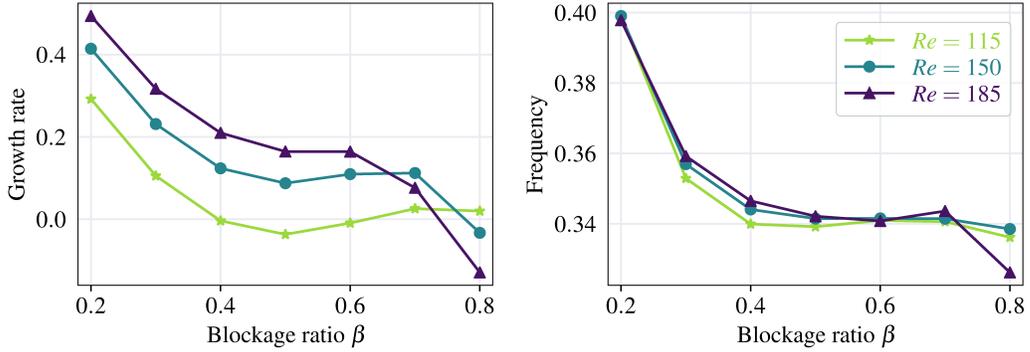}}
\caption{Growth rates and frequencies of SFD base flows of confinement cylinder wakes with different blockage ratios and Reynolds numbers.}
\label{fg_gr_omg}
\end{figure}

\begin{figure}
\centering
\subfigure[$Re=115$]{
\includegraphics[width=0.32\textwidth]{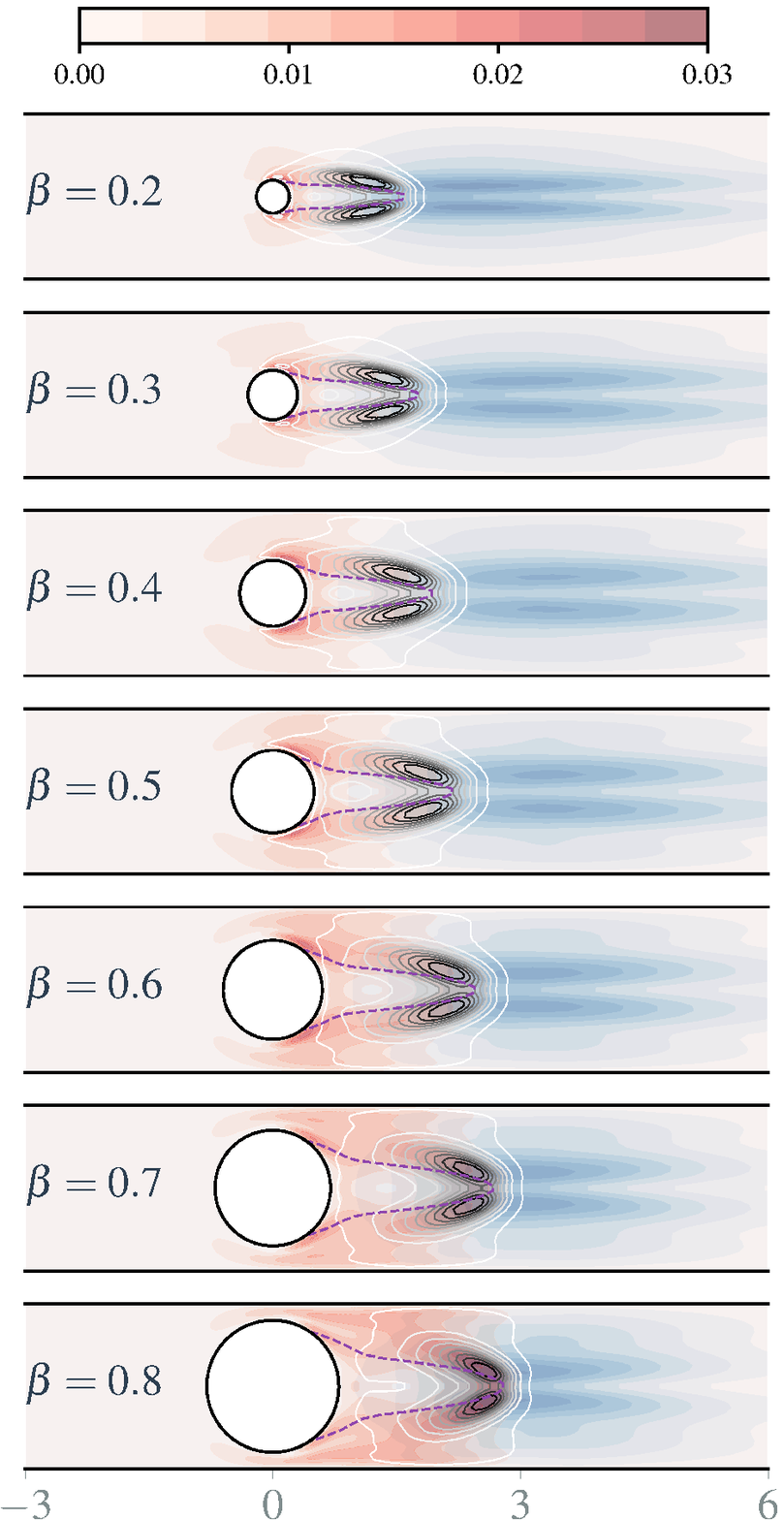}}
\subfigure[$Re=150$]{
\includegraphics[width=0.32\textwidth]{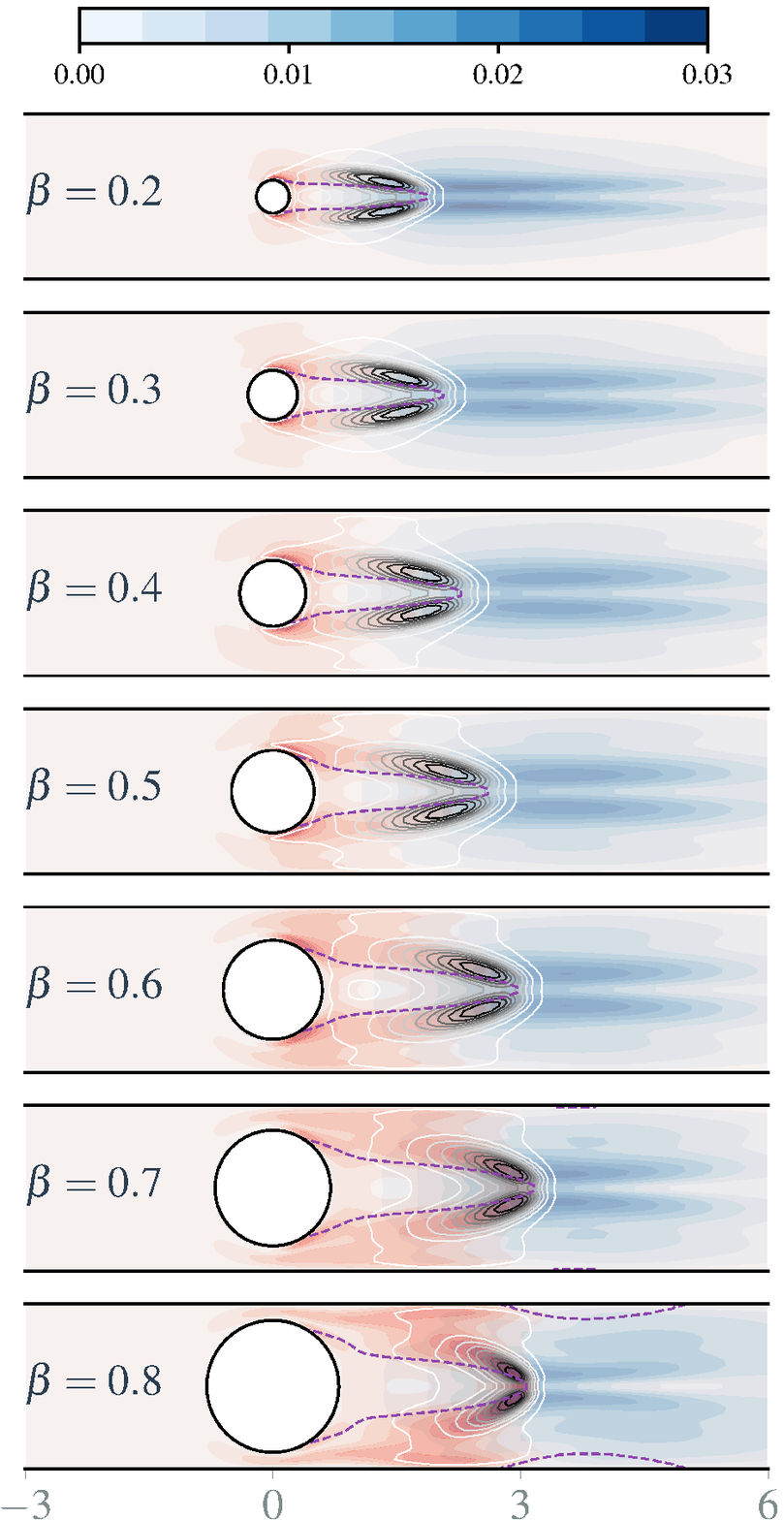}}
\subfigure[$Re=185$]{
\includegraphics[width=0.32\textwidth]{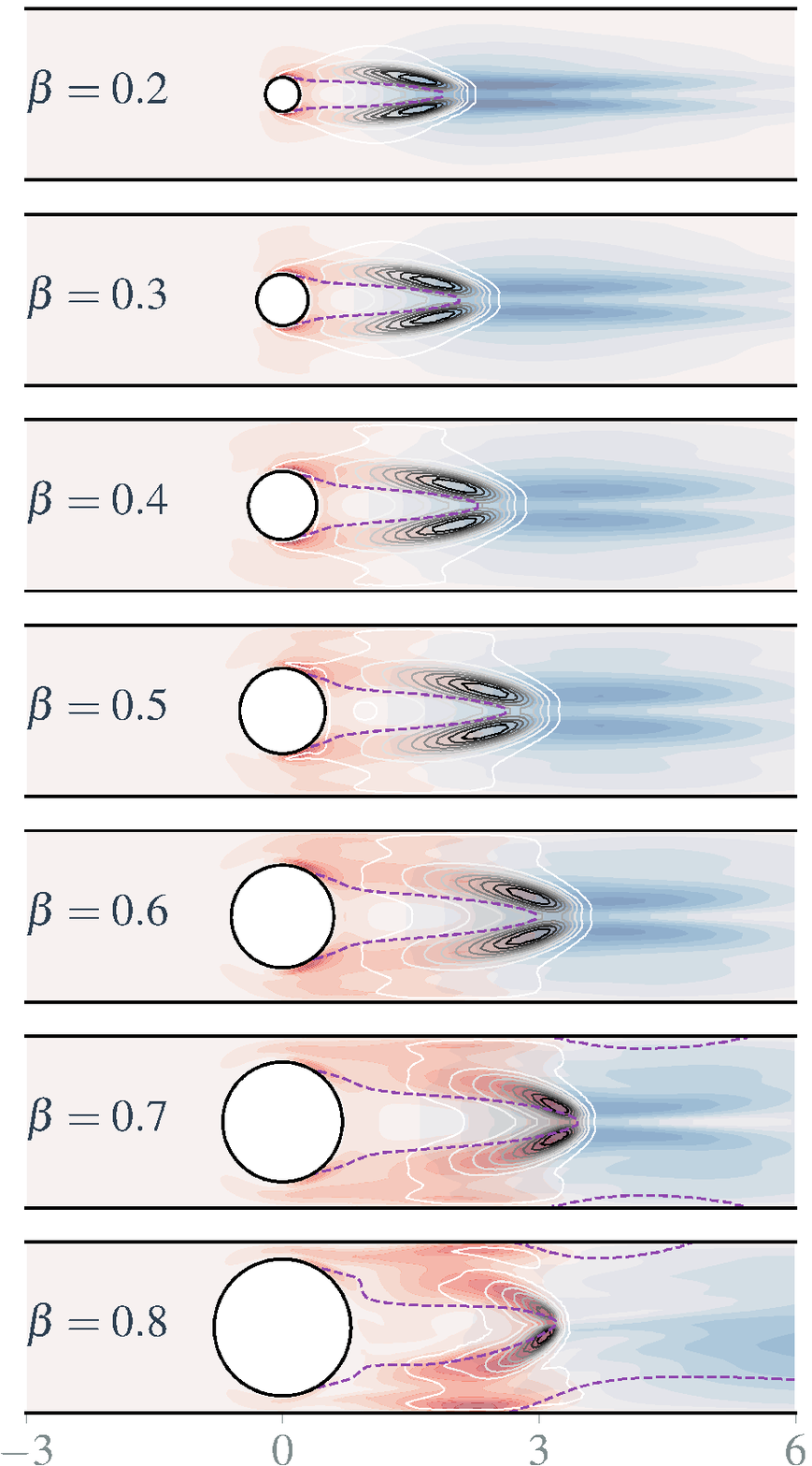}}
\caption{Structural sensitivity analysis of confined cylinder wakes with different blockage ratios and Reynolds numbers.
The adjoint mode, direct mode, and \textit{wavemaker} are shown by red, blue, and grey colormaps.
The recirculation zone is illustrated by the purple dash line. Note that only part of the domain is shown. 
}
\label{fg_wvmker}
\end{figure}

The above section details the frequency and growth rate of the dominant linear mode in the confined cylinder wake flow. Regarding the flow control and manipulation, much more useful information can be obtained by conducting a sensitivity analysis.  
Structural sensitivity reveals the most sensitive spatial part of the flow to perturbation; this region is traditionally dubbed as a \textit{wavemaker} region.
We investigate the influence of blockage ratios and Reynolds numbers on the \textit{wavemaker} region in this section.
Three Reynolds numbers, $Re=115, 150, 185$ are selected, covering subcritical and supercritical cases.
The growth rates and frequencies of the leading eigenvalues obtained in the global linear stability analysis are shown in figure~\ref{fg_gr_omg}.
When the blockage ratio $\beta$ is smaller than 0.5, increasing $\beta$ stabilizes the confined cylinder wake, and the leading eigenfrequency decreases simultaneously.
For the confined wake flow at $Re=115$, the vortex shedding is even fully suppressed (i.e., the flow becomes stable) when $0.4<\beta<0.6$.
Then, further increasing $\beta$ destabilizes the wake, which becomes unstable again when $0.6<\beta < 0.8$.
These results on the growth rate are consistent with those in figure \ref{fg_Recs}.
In confined cylinder wakes at $Re=150$, a similar stabilizing-destabilizing trend is observed when $0.4 < \beta < 0.6$ but the flow is always unstable with a positive growth rate.
A stabilizing effect is observed when $\beta > 0.7$ and the flow becomes stable when $\beta = 0.8$.
This is because the second critical Reynolds number ($Re_{c2}$) exists in the confined cylinder wakes with $\beta > 0.7$ and it decreases significantly with the increase of $\beta$, see figure~\ref{fg_Recs}. 
For wake flows at $Re=185$, the stabilizing effect for $\beta > 0.7$ is more significant and leads to a more stable flow when $\beta = 0.8$.
Despite the stability, the wake flow is in an asymmetric status (see figure \ref{fg_wvmker}c with $\beta=0.8$ below), and this phenomenon has been reported by~\cite{Sahin2004}.
On the other hand, the effects of $Re$ on the leading eigenfrequency are not significant when $0.4<\beta < 0.7$ as shown in the right panel of the figure.
When $\beta > 0.7$, the leading eigenfrequency of $Re=185$ decreases more significantly.

Next, we show the direct, adjoint modes, and \textit{wavemaker} region in the confined cylinder wake flow with different blockage ratios and Reynolds numbers in figure~\ref{fg_wvmker}.
The amplitudes of adjoint and direct modes for velocity are represented by the red and blue colormaps, respectively.
The \textit{wavemaker} region is shown by nine contour lines using the grey colormap,
which illustrates the sensitive domain with $\eta > 5\% \times \eta^\text{max}$,
where $\eta$ is the overlap function in structural sensitive analysis evaluated by Eq.~\ref{eq_wavemaker} and $\eta^\text{max}$ is the largest $\eta$ in the computational domain.
Similar to unconfined cylinder wakes~\citep{Giannetti2007,Marquet2008}, the \textit{wavemaker} region of the confined wake flow is located downstream the cylinder,
and thus to some extent, the upstream domain is less important in terms of flow sensitivity and needs not to be monitored in the suppression of wake vortex shedding.
Besides, when $\beta<0.7$, the cylinder wake flow with a larger blockage ratio (i.e., a stronger confinement effect) possesses a longer and wider recirculation zone, and correspondingly the \textit{wavemaker} region also expands and moves downstream. Regarding the length of the recirculation region, \cite{Chen1995} also reported longer recirculation zones as a function of $Re$, whereas the result of the elongated \textit{wavemaker} region with $\beta$ seems not to have been reported in the literature for the confined cylinder wake flow.
On the other hand, increasing the Reynolds number also leads to longer recirculation zones, and the \textit{wavemaker} region is pushed downstream (at least for the three $Re$ investigated here).
Regarding the flow control of the confined cylinder wake flows, the significance of these results is that in order to more efficiently suppress the vortex shedding, one should monitor the perturbations further downstream from the cylinder when $\beta$ or $Re$ increases because the most sensitive region (\textit{wavemaker}) is located further downstream.
Finally, for some wake flows with larger $\beta$, i.e., $Re=150, \beta=0.8$ or $Re=185, \beta=0.7, 0.8$, recirculation bubbles are developed on the confinement walls downstream the cylinder wake (similar to the results in figure \ref{fg_baseflows}),
which stabilizes the cylinder wake and leads to a second critical $Re$ as analysed in the previous section.

\section{Results: RL-based control of confined wake flows}
\label{secDiscussion}

In the above section, we have shown in detail the results of the stability and sensitivity analyses of the confined cylinder wake flow in order to obtain important flow information such as \textit{wavemaker} region.
In this section, we analyse the influences of key parameters in the RL algorithm on the control performance. 

\subsection{Vortex shedding suppression via reinforcement learning}
\label{sec_reward}
After characterising the linear instability and flow sensitivity of the confined cylinder wake flows, we adopt the deep RL to control the vortex shedding in this flow using the two synthetic jets on the cylinder. In the literature, \cite{Rabault2019} has studied RL-based control method to reduce the drag on the cylinder.

To begin with, we would like to first obtain some preliminary results by defining the reward as 
\begin{equation}
\label{eq_shedding_energy0}
  r_0=-\sum_i^{n_\text{node}} {((u^i)^2 + (v^i)^2)}
\end{equation}
in order to have a peek of what the controlled wake may become. $r_0$ is the negative value of a plain sum of the kinetic energy in the computational domain and $n_\text{node}$ is the number of grid nodes in DNS. Two points deserve to be mentioned that the kinetic energy will not decrease to zero as long as we feed the flow domain with an inflow and that the kinetic energy will decrease with the suppression of the vortex shedding. Thus, the RL algorithm will control the confined wake flow towards the lowest kinetic energy where the vortex shedding is weakest.
All the control investigations from Sec.~\ref{sec_reward} to Sec.~\ref{sec_neccesity} are based on 86 velocity probes covering the \textit{wavemaker} region (see probe distribution (b) in figure~\ref{fg_rough}).
The reason on this selection will be explained in Sec.~\ref{probe}.

\begin{figure}
\centering
\subfigure[Training process]{
\label{fg_RLcontrol_traing}
\includegraphics[width=0.485\textwidth]{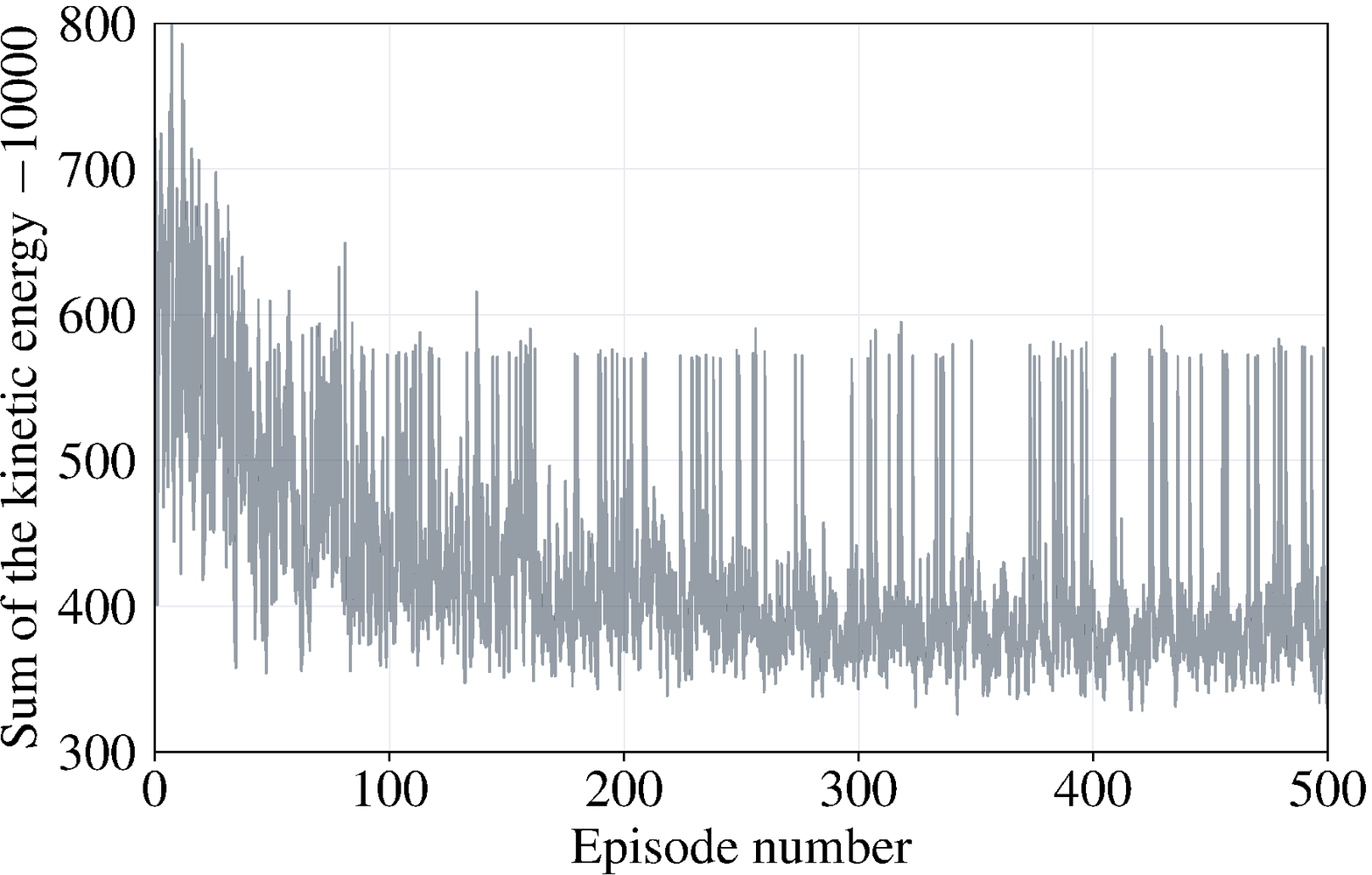}}
\subfigure[RL control performance]{
\label{fg_RLcontrol_performance}
\includegraphics[width=0.485\textwidth]{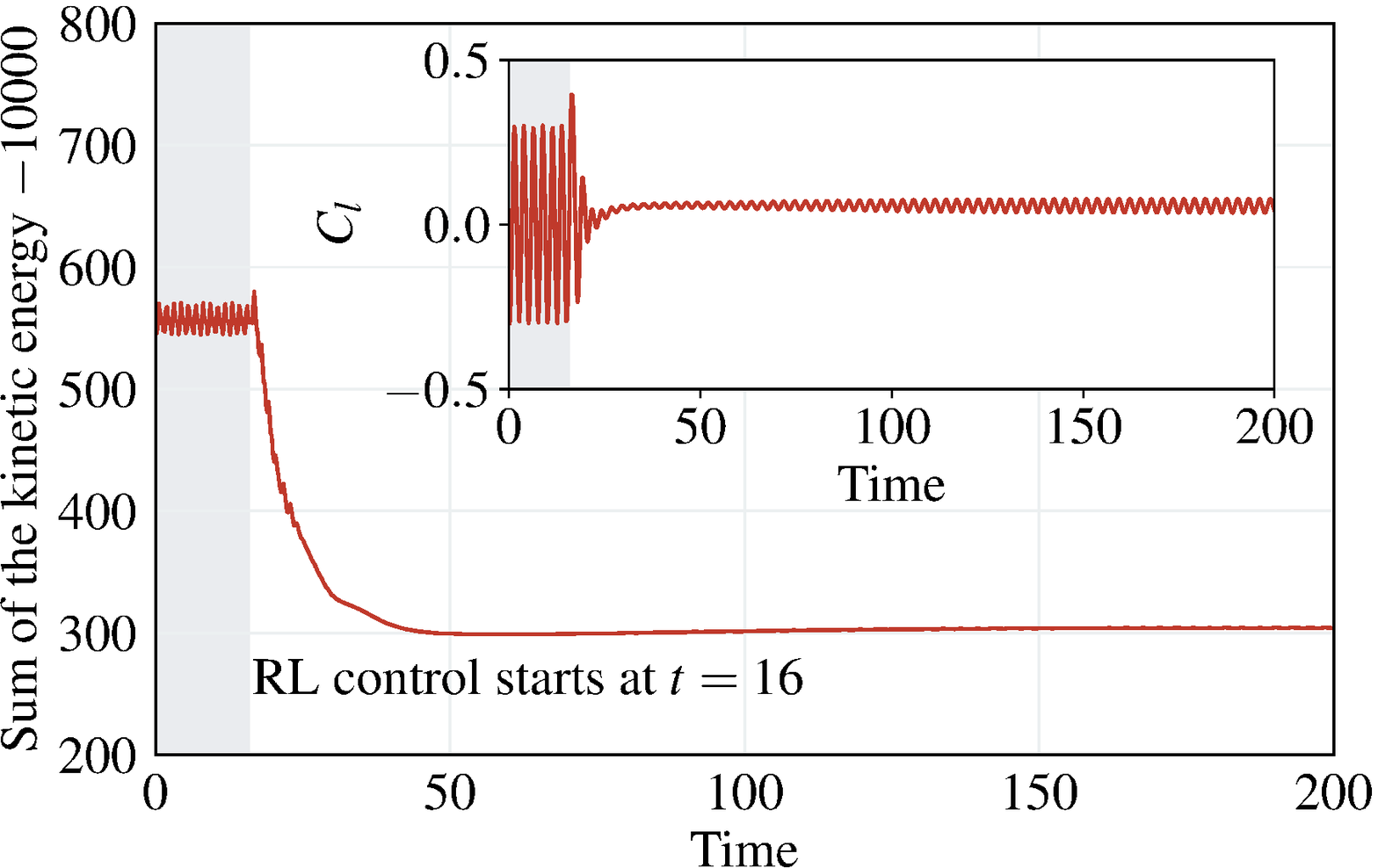}}
\caption{RL-based vortex shedding suppression control using the plain sum of the kinetic energy ($\beta=0.25$ and $Re=150$). In panel (b), the grey shade means that no control is applied until $t=16$. The reward function is Eq. \ref{eq_shedding_energy0}.  }
\label{fg_RLcontrol}
\end{figure}

\begin{figure}
  \centering
  \includegraphics[width=\textwidth]{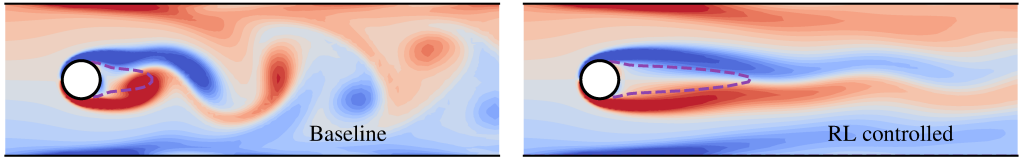}
  \caption{Vorticity of the baseline confined cylinder flow with $\beta=0.25$ at $Re=150$ and the controlled flow with the RL agent trained by the total energy.
  RL control using the kinetic energy damps the oscillation by increasing the recirculation zone.
  Nevertheless, the vortex shedding is not fully suppressed. The reward function is Eq. \ref{eq_shedding_energy0}. 
  }
  \label{fg_vorticity}
\end{figure}

Following~\cite{Rabault2019} , we use a random reset in the training process so that new episodes have a 20\% possibility to start from the given un-controlled initial condition and otherwise they start from the last state of the previous episode.
As shown in figure~\ref{fg_RLcontrol}a, the averaged energy decreases with the training of the RL agent, especially after 100 episodes.
The energy spikes of the learning curve result from the starting points of new episodes from the given initial condition.
After $400 \sim 500$ episodes, the RL training approximately converges. 
As shown in panel b of figure~\ref{fg_RLcontrol}, we test the RL-based control from $t=16$ to $t=200$ (the flow starts from the initial condition at $t=0$).
Although the testing control time is shorter than that in the fluidic pinball stabilization~\citep{CornejoMaceda2021}, the controlled flow has shown a convergence to a periodic state in this case.
The obtained control policy reduces the kinetic energy and the wake vortex shedding is suppressed (figure~\ref{fg_vorticity}).
At the beginning of the control (briefly after $t=16$), the energy firstly increases, and a slightly larger lift is observed consequently (see the inset in figure \ref{fg_RLcontrol}b). 
Then, the energy decreases exponentially, and the oscillations of the lift coefficient are also reduced before reaching a steady oscillation.
This result seems to be a common characteristics of the RL control applied to the wake flow in our case and will be seen later in our results.  A larger actuation has also been observed in the RL control of \cite{Tang2020} whose aim is to reduce the drag on the cylinder.
Eventually, the value of the lift coefficient converges to an oscillation with a small amplitude. 
Figure~\ref{fg_vorticity} shows the vorticity of the baseline and controlled flows.
RL control increases the recirculation zone when it suppresses the oscillation.
Nevertheless, slight vortex shedding downstream from the recirculation zone can still be observed, though the extent and amplitude are largely decreased.

\subsection{SFD base flow}
\label{sec_rewardSFD}
From figure \ref{fg_vorticity}, we see that when the confined wake flow is controlled, one salient feature is that the recirculation zone elongates. This is in fact a strong clue indicating that the controlled wake flow may converge to a flow that is close to the SFD base flow, because the recirculation zone in the latter case is relatively long (see figure \ref{fg_baseflows}). 

In order to testify this hypothesis, we use the fluctuation of the kinetic energy to monitor the strength of the vortex shedding,
which is computed by
\begin{equation}
\label{eq_shedding_energy}
  s_e = \sum_i^{n_\text{node}} { \left( (u^i - u^i_0)^2 + (v^i - v^i_0)^2 \right) }.
\end{equation}
The shedding wake flow ($\boldsymbol{u}$) can be decomposed as a sum of a reference part $\boldsymbol{u}_0=(u_0,v_0)$ and a shedding part $\boldsymbol{u}_s$, as in $\boldsymbol{u}  = \boldsymbol{u}_s + \boldsymbol{u}_0$. To suppress the vortex shedding, we can use $r = -s_e$ as the reward function in RL, which will minimize the fluctuation shedding energy. The idea is then to use the time-mean flow or the SFD base flow to evaluate the shedding energy in training RL control agents; that is, the mean flow $\boldsymbol{u}_b$ (obtained by time-averaging the periodic vortex shedding) and the SFD base flow $\bar{\boldsymbol{u}}$ (obtained by the SFD method) are used in Eq. \ref{eq_shedding_energy} for $\boldsymbol{u}_0$, respectively. If the RL control with the SFD base flow being used in the reward can yield $r\rightarrow 0$, then we can understand that the controlled wake flow indeed converges to the SFD base flow. The time-mean flow is also tested for a comparison.

\begin{figure}
\centering
\includegraphics[width=\textwidth]{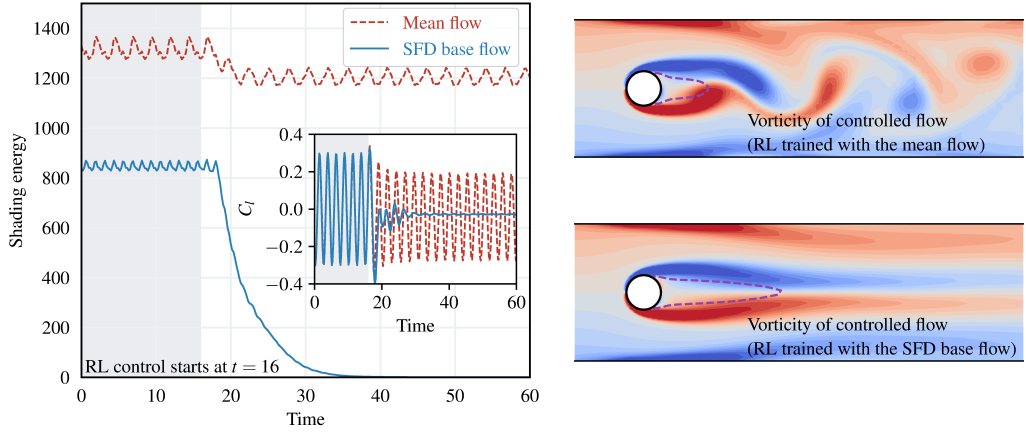}
\caption{RL-based vortex shedding suppression of the confined cylinder wake ($\beta=0.25$ and $Re=150$) with RL agents trained using different reference flows (Mean flow and SFD base flow).
It can be seen using the SFD base flow as the reference in the RL reward evaluation leads to an effective control with the vortex shedding fully suppressed. The reward function is Eq. \ref{eq_shedding_energy}.
}
\label{fg_RL_mean_vs_base}
\end{figure}

Figure~\ref{fg_RL_mean_vs_base} shows the RL control performances of the corresponding agents.
Using the mean flow in the shedding energy evaluation cannot effectively suppress the vortex shedding.
The result indicates that the time-averaged mean flow is modified to another status where the vortex shedding is slightly suppressed.
Thus, the mean flow is not the flow that the RL algorithm will lead the controlled wake flow to.
On the other hand, using the SFD base flow in the shedding energy evaluation (Eq.~\ref{eq_shedding_energy}) leads to a complete suppression.
The shedding energy is reduced to approximately zero, which means that the controlled wake is almost the same as the SFD base flow. 
The fluctuations of the lift coefficient are also almost fully suppressed, and no vortex shedding is observed in the vorticity contours.
It can be concluded that vortex shedding suppression (by RL) in the confined cylinder wake flow is realized by modifying the wake flow to be a status which is very close to the SFD base flow.
Interestingly, \citet{Flinois2015} found similar results in the adjoint-based optimal control of an unconfined cylinder wake that the controlled flow is the same as the SFD base flow.
Because its good performance, we will use the SFD base flow to evaluate the shedding energy for RL control in the following.

\begin{figure}
\centering
\includegraphics[width=\textwidth]{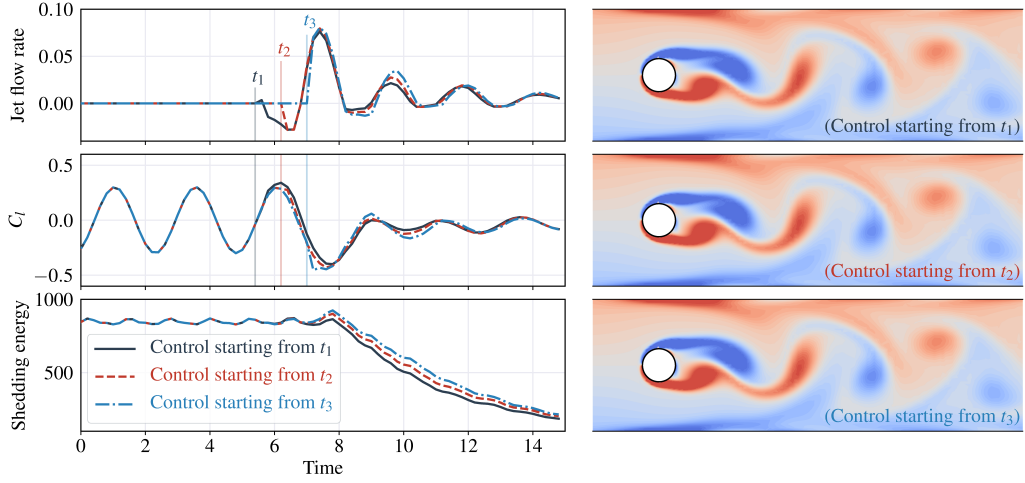}
\caption{Confined cylinder wakes ($\beta=0.25$ at $Re=150$) with the RL control starting at different time points ($t_1$, $t_2$, and $t_3$).
The right-hand sub-figures show the wake vorticities at $t=7.5$. 
It can be seen that, no matter when the control starts, RL always chooses to use large flow rates to offset the vorticity at the moment of the upper vortex rolling-up.}
\label{fg_RLctrl_diffphase}
\end{figure}

\subsection{Control starting time}
The above RL control starts at $t=16$, which is arbitrarily chosen. We next test the robustness of the RL control policy in terms of its application at other phase in the vortex-shedding period.
Figure~\ref{fg_RLctrl_diffphase} shows the control processes of the confined cylinder wake with the RL-based control starting from three different time stamps. 
The interval between adjacent time points is 0.8 unit time ($t_2 - t_1 = t_3 - t_2 = 0.8$), which is approximately $1/3$ period of the vortex shedding.
Among the three cases explored here, a general trend is that the trained RL always uses large positive jet flow rates ahead of the next minimum point of the lift coefficient, when the upper vortex rolls up.
Afterwards, relatively small jet rates are needed, and the shedding energy starts to decay.
Figure~\ref{fg_RLctrl_diffphase} also shows the vorticity fields at the moments with largest jet flow rates, which are exactly the moment of the upper vortex rolling-up.
The large blowing of the upper jet produces positive vorticity to offset the rolling-up vorticity, which can be considered to be destructive to the vortex development.
This suggests that using large flow rates when the vortex rolls up is an important step in suppressing the cylinder wake fluctuations.
The trained RL agent is adaptive to the shedding phase and captures the ``right'' timing.
Due to the symmetry of the problem at hand, it can be understood that there is a similar RL control policy that uses significant blowing of the lower jet to suppress the rolling-up of the lower vortex.

\subsection{Necessity of a persistent oscillating control}
\label{sec_neccesity}
\begin{figure}
\centering
\includegraphics[width=\textwidth]{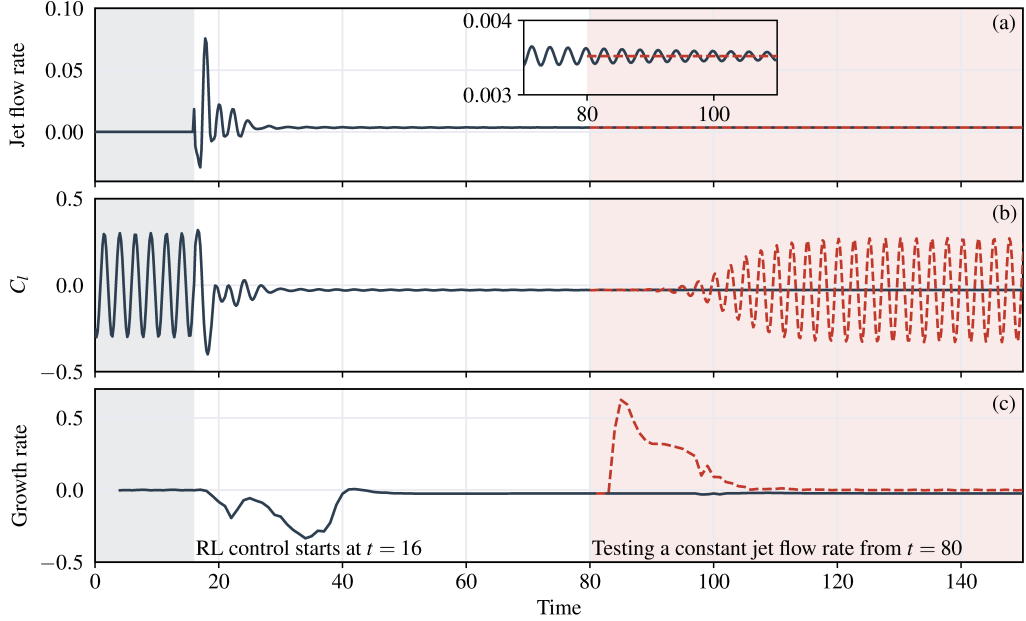}
\caption{DMD analysis of the RL-based control process.
The gray shade is the confined cylinder wake without control, and the RL-based control starts at $t=16$.
We test a constant jet flow rate in the controlled cylinder wake from $t=80$, which is shown as the red shade.
The black solid lines show results of RL-based control and the red lines are results with the control switched to the constant flow rate.
}
\label{fg_DMDcontrolled}
\end{figure}

\begin{figure}
\centering
\includegraphics[width=\textwidth]{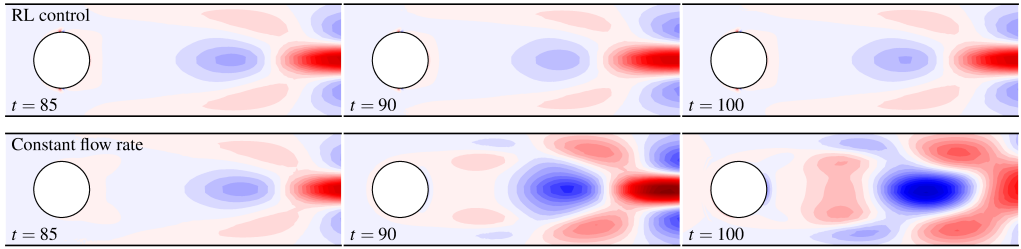}
\caption{Vorticities of the leading DMD modes in the controlled flows ($\beta=0.25$ at $Re=150$). 
Merely part of the computational domain is shown for clarity. 
After the vortex shedding is damped, using a similar but constant jet flow rate soon triggers flow instability and gradually leads to vortex shedding in the cylinder wake.
}
\label{fg_leadingmodes}
\end{figure}

According to the linear stability analysis, the SFD base flow is unstable; the RL agent maintains this unstable status by actively modifying the jet flow rate in real time.
In order to probe how the stability property of the flow changes with time, we will use DMD ($N_\text{snapshot}=20$ and $\Delta t = 0.2$) to monitor the status of the flow stability/instability in the control process.
In additional tests (not shown), we have used more DMD snapshots and found that the results agreed well with those to be presented in the following.
In figure \ref{fg_DMDcontrolled}, the control starts at $t=16$, and in the beginning, again, relatively large jet flow rates are imposed by the RL agent.
Then, as the lift coefficient is reduced, the flow rate of the synthetic jets decreases to a small value with a very small oscillating amplitude (see the inset in panel a).
Even though the amplitude of the fluctuation is insignificant, we find that this fluctuation in the flow rate is vital to the successful continuous suppression of vortex shedding (we note in passing that the jet flow rate does not oscillating exactly around 0 because the overall effect of blowing/suction will introduce a small degree of asymmetry with respect to the channel centreline in the flow).
We perform a numerical experiment in figure \ref{fg_DMDcontrolled} to impose a sudden modification to the jet flow rate after $t=80$. If nothing is changed in the RL control, when $t>80$, the controlled wake flow is stable with the slightly fluctuated flow rates (the black lines).
However, using a similar but constant jet flow rate ($u=0.00352$, see the red dashed line in panel $a$) at $t=80$ soon triggers flow instability (panel $c$) and gradually leads to vortex shedding in the cylinder wake (panel $b$).
Figure~\ref{fg_leadingmodes} shows the leading DMD modes of the two flows from $t=85$ to $t=100$.
With the active jet flow rate fluctuation, significant vorticity is observed around the jets in the leading mode of the RL controlled flow.
On the other hand, the leading DMD mode almost remains unchanged in the (successful) controlled process from $t=85$ to $t=100$.
But if a constant jet flow rate is forced, the vorticity around the jets of the leading mode vanishes. At the same time, the leading mode becomes unstable and gradually evolves to a state with much stronger vorticities downstream the cylinder. 
Finally, the flow develops to a saturated periodic vortex shedding state, which has a similar $C_l$ magnitude to the uncontrolled flow (see panel $b$ in figure \ref{fg_DMDcontrolled}).
To sum up, in the vortex shedding suppression of cylinder wake, RL-based policy tends to spend large energy at the start of the control in modifying the mean flow structure.
Then, much less effort is required to maintain the stabilised flow.
A similar tendency is found in the RL-based drag reduction control of a confined cylinder~\citep{Rabault2019}.
Nevertheless, a persistent active control using the RL agent, even though its oscillating amplitude is small, is required to suppress the vortex shedding in the cylinder wake. This is because the controlled flow is almost identical to the SFD base flow, which is unstable.
Thus, when the persistent control is suppressed, the stabilised flow will become unstable and the vortex starts shedding again, as demonstrated in our numerical experiment above. 

\begin{figure}
\centering
\includegraphics[width=\textwidth]{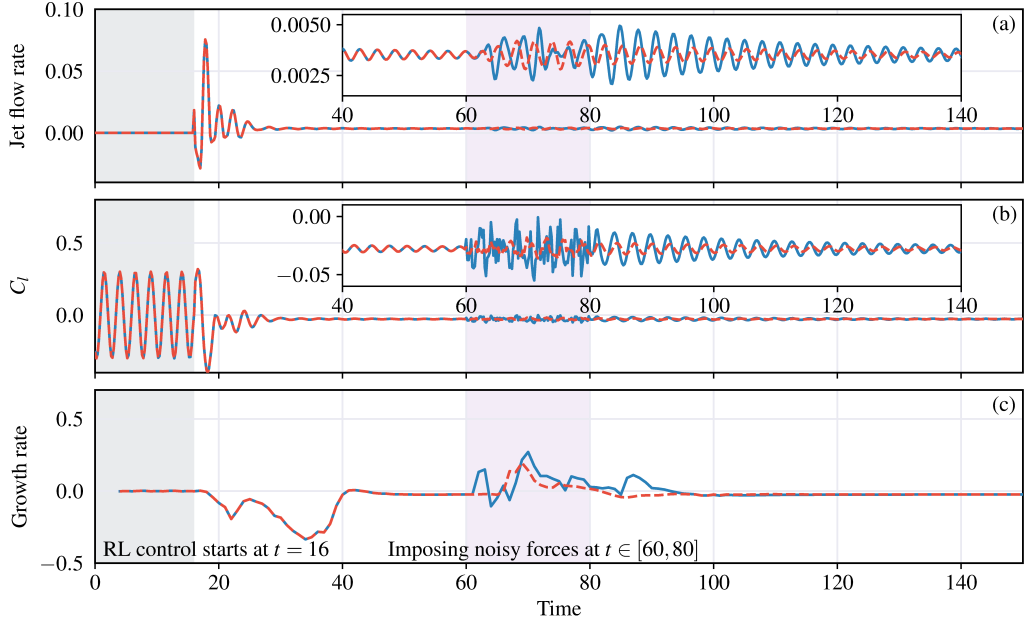}
\caption{DMD analysis of the RL-based control process with noisy forces in the inflow domain, which is imposed at $t \in [60,80]$.
We test two force magnitudes ($A=2.0$ and $A=10.0$), which is shown as the red and blue lines, respectively.
The controlled wake flows are not developed to significant vortex shedding and get stabilised again after the noisy force is removed.
}
\label{fg_noisyInflow}
\end{figure}

In real-world applications, the control system always faces uncertain noise. 
\cite{Tang2020} and \cite{paris2021robust} have shown that the RL-based control can be robust to the Reynolds-number variation and systematic noise.
There is another uncertainty which is the incoming disturbances penetrating into the inlet domain.
To study this influence, we place an external forcing term $\boldsymbol{F}(x,y) \gamma(t)$ following \citet{Herv2012}.
$\gamma(t)$ is a random scaling factor of standard deviation $\sigma_{\gamma} = 0.1$.
$\boldsymbol{F}(x,y)$ is the spatial structure with a Gaussian shape
\begin{equation}
\label{eq_extForce}
\boldsymbol{F}(x,y) = A \exp{\left( \frac{-(x-x_0)^2}{2 \sigma_{F}^2} \right)} \exp{\left( \frac{-(x-x_0)^2}{2 \sigma_{F}^2} \right)} (1,1)^T,
\end{equation}
where $\sigma_{F} = 0.1$ and the spatial center is $x_0=-1.5$ and $y_0=0$. The magnitude of the external force can be modified by changing $A$, and two values ($A=2.0$ and $A=10.0$) are studied.
Figure~\ref{fg_noisyInflow} shows the RL-based control of the cylinder wake flow with noisy forces imposed at $t \in [60,80]$.
To stabilize the flow, jet flow rates controlled by the RL agent are increased slightly, and the increase magnitude is positively correlated with $A$.
Although the noisy spatial forces trigger instabilities in controlled wake flow, the wake does not develop significant vortex shedding.
After the noisy force is removed ($t > 80$), the wake gets fully stabilised again after ~20 unit time.
This numerical experiment shows that RL-based control is robust to spatial noise from the inflow and it is vital to sustain a persistent RL-based control.

\subsection{Stability-enhanced reward}
\begin{figure}
  \centering
  \includegraphics[width=\textwidth]{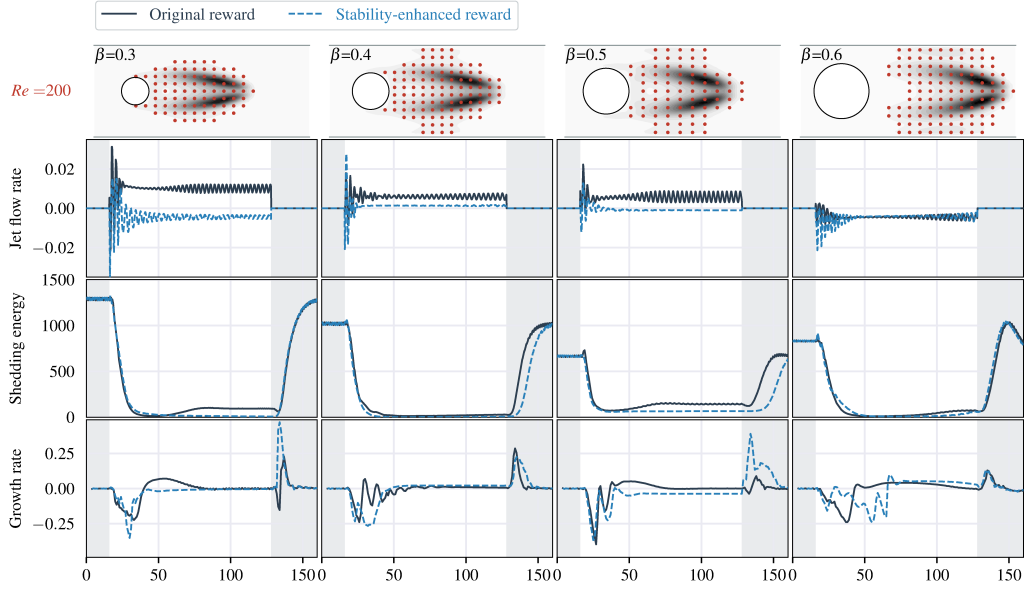}
  \caption{RL control performance of cylinder wakes with different $\beta$ at $Re=200$. The black region indicate the \textit{wavemaker} region and the probes are placed to cover it (see the discussion in Sec.~\ref{probe}).
  The stability-enhanced reward reduces unstable instants in the controlled wake flows.
  From $t=128$, the RL-based control is shut down (shown as the red shade), and the controlled wake flows re-develop to significant vortex shedding, which shows the necessity of a persistent control.
  }
  \label{fg_diffprobesMoreRe200}
\end{figure}

Next, we discuss how the reward in RL can be changed to incorporate some information on flow stability/instability to improve the control performance. 
Increasing the Reynolds number destabilizes the confined wake flows, especially for cylinders with $\beta \le 0.5$.
In the RL-based drag reduction control of a confined cylinder wake with $\beta \sim 0.25$, \cite{Tang2020} found that small drag oscillations still existed in the controlled flow at $Re=200$ although four synthetic jets were used; and with a further increase of $Re$, although a significant reduction of the averaged drag was achieved, the RL agent cannot find a fully stabilised control strategy to completely suppress oscillations in the drag coefficient.
This means that the strong instability with the increase of $Re$ in confined cylinder wakes brings significant challenges to RL-based control, and thus full suppression of the vortex shedding is difficult. 
We increase the Reynolds number to $Re=200$ (note the difference in the definitions of $Re$ in our work ($Re= U_{\text{max}}D/ \nu$) and their works ($Re= \bar{U} D/ \nu$); $Re=200$ in our work is corresponding to $Re=400/3$ in their work) and investigate its influence on vortex shedding suppression with different $\beta$.
As shown in figure~\ref{fg_diffprobesMoreRe200}, the increase of $Re$ further pushes the \textit{wavemaker} domains a bit downstream of the cylinder.
We place the probes for RL control to cover the \textit{wavemaker} region (to be discussed in Sec. \ref{probe} below) and train RL agents for vortex suppression at $Re=200$ using the original reward in Eq. \ref{eq_shedding_energy}.
It can be seen that not all the controlled wake flows are fully stabilised due to the increase of $Re$.

We used DMD ($N_\text{snapshot}=20$ and $\Delta t = 0.2$) to analyse the controlled flows and found that the growth rate may flip to be positive even after the shedding energy is significantly reduced, which leads to flow instabilities and a subsequent increase of shedding energies.
This unstable control is not preferable in practice.
In order to obtain an effective control policy, we use a stability-enhanced reward function for RL, which embeds the information of flow stability/instability in the original reward:
\begin{equation}
  \label{eq_reward_growthrate}
  \text{Reward} =-s_e {e^g},
 \end{equation}
where $s_e$ is the shedding energy defined in Eq.~\ref{eq_shedding_energy}, and $g$ is the largest growth rate of the flow, which is evaluated by DMD.
For a stable flow, the growth rate is smaller than zero and $e^g < 1.0$, which means a larger reward is given to motivate such control policies.
For a neutral flow, the growth rate is zero, and Eq.~\ref{eq_reward_growthrate} is equivalent to using the original reward defined by the shedding energy.
For unstable flows, the growth rate is larger than zero.
In this circumstance, the stability-enhanced term $e^g > 1.0$, and the reward decreases correspondingly.
With such a reward function, the trained RL agent will come up with a control policy that compensates for the decrease of the reward due to flow instability. This mitigates the adverse effect of flow instability in the control.
We train RL agents using the stability-enhanced reward for vortex suppression with different $\beta$ at $Re=200$.
As shown in figure~\ref{fg_diffprobesMoreRe200}, using the stability-enhanced reward, the instants when the growth rate is larger than zero become fewer in the controlled wake flows. 
Thus, instabilities in the controlled flows are significantly damped.
Such an instability-abated effect is of great value in practical applications such as prevention of aerodynamic buffeting~\citep{Gao2017} and aeroelastic flutter~\citep{Jonsson2019}.
In most cases (figure~\ref{fg_diffprobesMoreRe200}), the shedding energy with the stability-enhanced reward is reduced to a lower value that can be maintained for a long run, and, in these circumstances, the amplitude of the control actions is decreased as well.
This implies that it is favourable to maintain the controlled cylinder wake flow close to the base flow because it costs lower control energy, which is a beneficial side effect of the proposed stability-enhanced reward.
Nevertheless, there is still no guarantee to get a fully stabilised wake flow by using the proposed stability-enhanced reward (see $t \in [100, 128]$ in the case of $\beta=0.6$).
We also investigate the destabilizing process of the controlled flow with the shut-down of control actions (shown as the red shades in figure~\ref{fg_diffprobesMoreRe200}).
The results show that a continuous control is necessary; otherwise, the flow will become unstable.

We provide some discussions on these results in terms of flow physics. In the literature of 2D cylindrical wake flow, the shift mode has been found to be indispensable in a successful reduced-order model (ROM) of the transient dynamics in this flow \citep{Noack2003}. Inclusion of this mode in a ROM realises a mean field correction, pointing from the unstable steady flow to the time-averaged mean flow. Our results on the RL of the confined wake flow can be understood in a similar manner. The controlled flow is the SFD base flow, which is unstable and tends to evolve to the periodic flow once the control is turned off (see figure \ref{fg_DMDcontrolled}). Thus, when we improve the RL agent by telling it how to abate the instability, the control can be more effective and efficient, as shown in our tests (see figure \ref{fg_diffprobesMoreRe200}). Admittedly, our stability-enhanced reward is still crude as we simply incorporate the information of the first DMD mode in the new reward. 
In answering a comment by one of the reviewers, we realise that the analysis and RL control can be conducted more delicately by abating the flow instability in the direction of the unstable base flow drifting to the time-mean flow (currently we do not know how much the abatement effect of using the first DMD mode can contribute in this direction). This may lead to a more efficient RL control. More research is required to explore the new design of reward function in RL in the future.

\subsection{A heuristic strategy of probe placement}
\label{probe}

\begin{figure}
  \centering
  \includegraphics[width=\textwidth]{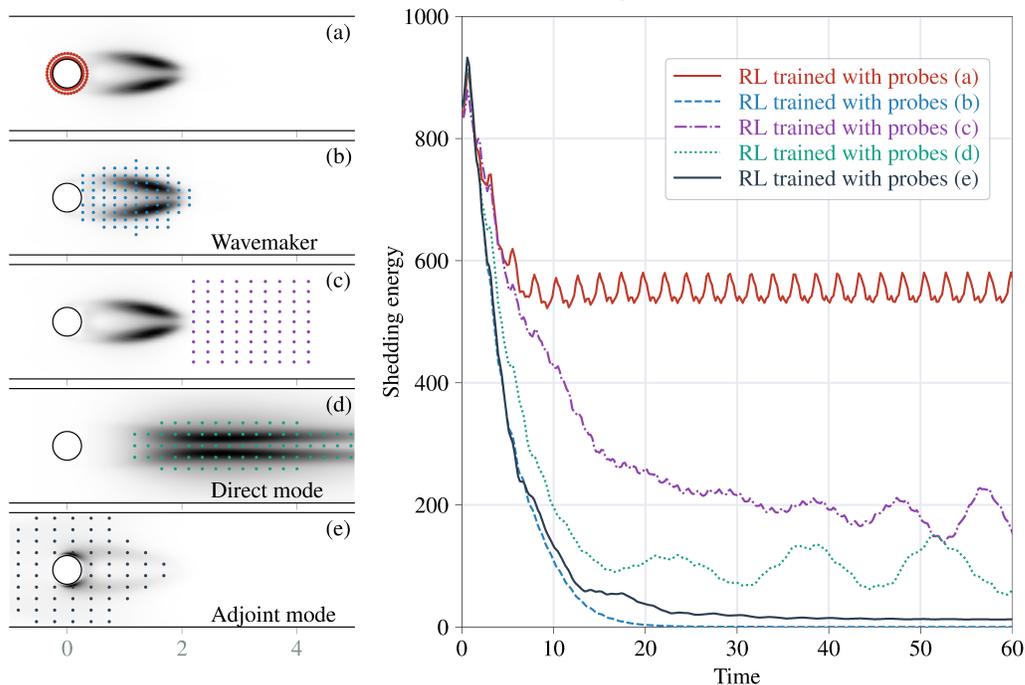}
  \caption{Probes covering the \textit{wavemaker} zone lead to the best control of the confined cylinder wake ($\beta=0.25$ at $Re=150$). }
  \label{fg_rough}
\end{figure}

The reinforcement learning agent observes environment changes via the probes.
In preparing this work, we found that \citet{paris2021robust} studied the probe placement by modifying the architecture of RL-based control by adding a stochastic gated input layer between the states and agent. 
The modified RL can select an optimal subset from the initial probe sensors. 
However, a good control performance in this case may be largely dependent on the initial placement of the probes. It may be the case that the initial placement of the probes is not able to encompass the global optimal solution of the control. To achieve an effective RL-based control, probes should be placed in domains that can convey all the essential changes in the flow. In the following, we take a heuristic approach to evaluate the placement strategy of the probes by harnessing the results of the structure sensitivity. 

\begin{figure}
  \centering
  \includegraphics[width=\textwidth]{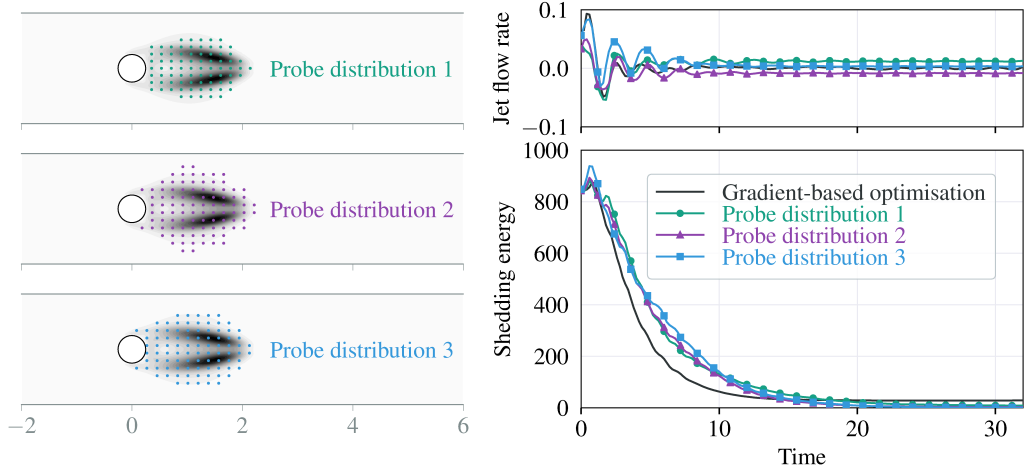}
  \caption{Even with slight differences in the placement, probes covering the \textit{wavemaker} zone can always lead to effective and stable RL control of the confined cylinder wake ($\beta=0.25$ at $Re=150$).
  The optimal control policy solved by the gradient-based optimisation achieves better short-term performance, and RL agents lead to a better performance in the long term.
  }
  \label{fg_diffprobes}
\end{figure}

\begin{figure}
  \centering
  \includegraphics[width=\textwidth]{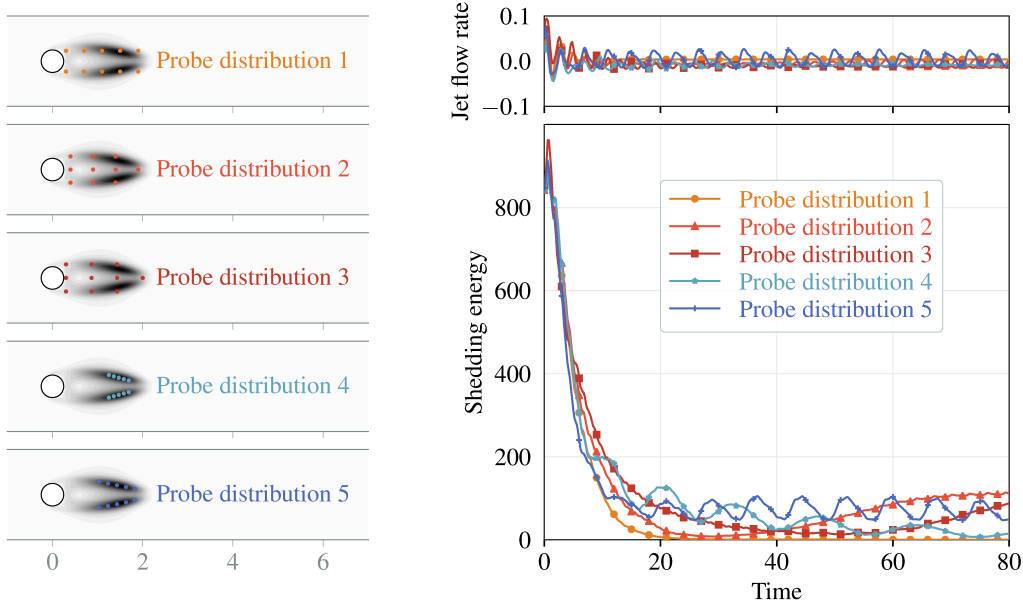}
  \caption{RL control of the confined cylinder wake ($\beta=0.25$ at $Re=150$) using ten probes.
  Different distributions of probes lead to a significant divergence in the control performance.
  }
  \label{fg_insuffprobes}
\end{figure}

As the most sensitive region to flow perturbations, the \textit{wavemaker} region may be the ideal monitoring position for the probe placement.
\citet{Strykowski1990} showed that the \textit{wavemaker} is similar to the region where one can place a small control cylinder to suppress the vortex shedding. 
We hypothesize that the most effective probes should be placed to cover (at least partially) the \textit{wavemaker} domain in RL-based control of vortex suppression in confined cylinder wakes.
We place the probes according to the \textit{wavemaker} region calculated using the SFD base flow. In the following, three sets of tests have been conducted as shown in figures \ref{fg_rough},\ref{fg_diffprobes},\ref{fg_insuffprobes}, which are explained below.

First, in figure \ref{fg_rough}, we would like to demonstrate that the probes should be placed to cover the \textit{wavemaker} region. In order to verify this, we consider three kinds of probes: 1) probes upstream of the \textit{wavemaker}, 2) probes inside the \textit{wavemaker}, 3) probes downstream of the \textit{wavemaker}.
As shown in figure~\ref{fg_rough}, placing the probes upstream the \textit{wavemaker} significantly reduces control effectiveness (even though they are very close to the cylinder);
while placing the probes downstream of the \textit{wavemaker} cannot train a stable control policy either, even though the RL performance is better than the previous case.
Probes placed inside the \textit{wavemaker} have an effective and stable RL control.
Since the \textit{wavemaker} is obtained by overlapping the direct mode and adjoint mode, we further investigate the control effectiveness with probes being placed to cover the direct mode and adjoint mode, respectively.
As shown in figure~\ref{fg_rough}, the RL-based control with probes covering the direct mode is ineffective, while the control with probes covering the adjoint mode is effective and the performance is almost close to those covering the \textit{wavemaker}.
This may be due to the fact that the adjoint mode covers most of the \textit{wavemaker} domain while the direct mode is located much downstream.
Although the direct mode describes the spatial structure of vortex shedding and most residual non-stationarities in the controlled wake flows are located in the region of the direct mode, it is more effective for RL-based control to monitor flow changes in the core domain of instability.

Next, in figure~\ref{fg_diffprobes}, we would like to demonstrate that a small difference in the probe distribution will not affect the general good function of RL as long as the main part of the \textit{wavemaker} is covered, confirming the robustness of its performance. All the control policies also display the similar features found in the last section; that is, large flow rates of the synthetic jets are used in the beginning to modify the mean flow structure, and much smaller time-varying jet flow rates are required to maintain the status.
In the same figure, we have also added a result of a gradient-based optimisation method in reducing the shedding energy. This consideration is to understand how the RL-based policy is compared to other control methods. In the gradient-based optimisation method, the flow rates at different time steps are discretised as independent design variables with a control horizon of 32 time units.
The objective function of the control design optimisation is the total shedding energy in the control time-horizon.
The gradient of the objective function with respect to the flow rates is solved by using the finite difference method. 
More details are provided in Appendix~\ref{AppendixC}.
It can be seen in figure~\ref{fg_diffprobes} that placing the probes in the \textit{wavemaker} region achieves a control performance that is close to that solved by the gradient-based optimisation.
The control policy found by gradient-based optimisation leads to more reductions of the shedding energy in the short-term ($t\in [5,15]$), since this strategy can achieve the smallest total shedding energy in the chosen control horizon (meaning that the controlled solution based on the gradient-based optimisation method is optimal in the sense that the total energy with $t\in[0,32]$ is minimum).
Nevertheless, it cannot be guaranteed that other solution may exist that has locally smaller shedding energies than the policy found by the gradient-based optimisation. For example, our RL-based controls have slightly better long-term performances (for instance, look at the results at $t=30$ in the figure), which can be more preferable in practice. 

The \textit{wavemaker} domain of the confined cylinder wake changes with the blockage ratio or the Reynolds number.
We investigate the influence of different blockage ratios on RL-based control performance at $Re = 150$. 
As shown in figure~\ref{fg_diffprobesMoreRe150} of Appendix \ref{AppendixD}, increasing the blockage ratio pushes the \textit{wavemaker} to the downstream region, as we have discussed above. 
We place probes for RL control of confined wakes with different blockage ratios at the corresponding \textit{wavemaker} regions. 
With the RL-based control, the shedding energies are all reduced to a small value. 
This means that, regardless of the blockage ratio, as long as the probes are placed to cover the \textit{wavemaker} region, the RL-control policy is able to effectively control the confined flow to damp the vortex shedding (at least for the Reynolds number investigated, $Re = 150$). 
In practice, if the probes are fixed in a series of control tests with different parameters (such as $\beta$ or $Re$), the optimal placement of the probes should be determined to be suitable for all the parameters (the trend of how the flow properties change with these flow parameters have been analysed in the previous sections). 

Finally, we consider the cases when the number of probes is insufficient.
We choose to train the RL agent using ten probes. The probe distributions and control performances are shown in figure~\ref{fg_insuffprobes}.
Distribution $1\sim3$ are evenly placed in the \textit{wavemaker} region, and distribution $4\sim5$ are placed in most sensitive part (black).
We can see that even though all the 5 placements successfully reduce the shedding energy, difference exists in their performance. The best performance belongs to Distribution 1, where the shedding energy has been abated to almost zero when $t>20$. This result indicates that Distribution 1 should be close to the global optimal solution in the RL (in which the shedding energy is zero). On the other hand, Distributions 2,3, even though they are also trying to cover the \textit{wavemaker} region as fully as possible, somehow perform poorly due to the insufficient number of probes. Note that, compared to figure \ref{fg_diffprobes}, the results of Distributions 2,3 should be considered worse. Finally, when clustering all the available probes in the most sensitive region (black colour) as in Distributions 4,5, the RL control does not necessarily yield good performance. All these results indicate that when insufficient probes are used, the performance of the RL control is scattered depending on the placement strategy of the probes. Our heuristic approach indicate that Distribution 1 is the best choice; however, this fortuitous result may not be carried over to other situations. Based on this result, future works can consider coupling the proposed criterion of probe placement (to cover the \textit{wavemaker} region) with an optimisation method with the former providing a general good choice of the initial placement of the probes and the latter fine-tuning the selection of the probes by the optimisation of the probes, exemplifying the combination of the priori knowledge on flow physics and the power of RL algorithm.
For example, the modified RL developed by~\citet{paris2021robust} can be used to select the optimal subset from the  probes which are initially covering or not covering the \textit{wavemaker} region (and we hypothesise that the probes covering the \textit{wavemaker} region will be selected when vortex shedding is to be abated). 

\section{Conclusions}
\label{secConclusion}

In this work, we have studied the linear stability and flow sensitivity of confined wake flows in a range of $\beta$ and $Re(\le 200)$. The main objective of the work is to understand further the reinforcement learning (RL) algorithm in controlling a complex flow system and showcase how some information of the flow physics can help us in designing and applying the RL in the flow control (more specifically, to suppress the vortex shedding by actively changing the flow rates of two synthetic jets on the cylinder). 

\subsection{Flow stability and sensitivity of confined wake flows}
In the stability study of the confined cylinder wakes, the neutral curve in the $Re-\beta$ plane is determined.
In a range of $Re \in [0,200]$, the confined cylinder wake with $\beta<0.7$ has one critical $Re$, where the wake becomes unstable beyond this point and the vortex shedding starts.
For confined cylinder wakes with $\beta>0.7$, other than the main recirculation bubble downstream the cylinder, two additional recirculation bubbles may develop close to the confinement walls further downstream the cylinder with the increasing effect of confinement.
This is similar to the observations by~\cite{Sahin2004}, and the recirculation bubbles on the confinement walls become larger with the increase of $Re$. This is associated with a second critical point, after which the vortex shedding is suppressed, for confined cylinder wakes with $\beta>0.7$.
Vortex shedding of confined cylinder wakes with different $\beta$ and $Re$ is studied using the global linear stability analyses.
Similar to the findings by~\cite{Maurel1995,Noack2003} in unconfined cylinder wakes, we find that the oscillating wakes of confined cylinders saturates when the time-averaged flow is marginally stable.
The frequencies solved by nonlinear DNS agree well with the results of the global linear analysis based on the time-mean flow, rather than the SFD base flow.
Thus the mean flow can provide a good profile to predict the shedding frequency of confined cylinder wakes.
The relationship between the SFD base flow and the saturated mean flow is studied by using global linear stability and DMD.
Similar to the unconfined flow past a cylinder \cite{Barkley2006}, the evolution of the unstable SFD base flow in the confined cylinder wake is a nonlinear saturation of oscillations, and the selection of vortex shedding amplitude and frequency is based on the marginal stability of the mean flow. 

The \textit{wavemaker} regions of confined cylinder wake flows with various $\beta$ and $Re$ are obtained by performing structural sensitivity analyses.
Similar to unconfined cylinder wakes~\citep{Giannetti2007,Marquet2008}, the \textit{wavemaker} region of the confined wake flow is located downstream the cylinder, which means the downstream domain is more important in terms of flow sensitivity and should be monitored in the suppression of wake vortex shedding.
We found that, with the increase of $\beta$, a longer and wider recirculation zone is developed and the \textit{wavemaker} region expands correspondingly and moves downstream.
Increasing $Re$ also leads to longer recirculation zones and the \textit{wavemaker} region is pushed downstream.
The results imply that, in order to efficiently suppress the vortex shedding in confined cylinder wakes, perturbations further downstream from the cylinder should be monitored when $\beta$ or $Re$ increases because the most sensitive region (\textit{wavemaker}) is located further downstream.

\subsection{RL-based control of confined wake flows}
In the second part of this work, we used RL-based control in order to suppress the vortex shedding in the cylinder wake.
It is found that using the sum of kinetic energy as the reward function can suppress the oscillation to some extent but did not fully damp the shedding.
We define a reward function based on the shedding energy relative to the SFD base flow, and this reward gives rise to a more effective RL control.
With the shedding energy being reduced to approximately zero, it can be concluded that vortex shedding suppression of the confined cylinder wake flow is realized by modifying the confined wake flow to a status similar to the SFD base flow.
Similar results have been observed in the adjoint-based optimal control of an unconfined cylinder wake in \cite{Flinois2015}.
We found that the RL-based control tends to spend large energy in the beginning to fast modify the mean flow structure.
No matter when the control starts, the trained RL agent is adaptive to the shedding phase and can capture the ``right'' timing, i.e., at the moment of the vortex rolling-up, to excite large flow rates.
Afterwards, much less effort with significantly small jet flow rates is required to maintain the stabilised flow.
Nevertheless, consistent active control using the RL agent is still necessary; otherwise, the stabilised flow will become unstable and the vortex starts shedding again. This is because the controlled flow is the unstable SFD base flow.

For confined cylinder wakes with $\beta \le 0.5$, increasing $Re$ brings more challenges to the vortex shedding suppression due to the rise of flow instability (from the modal stability analysis, we know that for smaller $\beta\le0.5$, the flow becomes more unstable when $Re$ increases).
In this circumstance, we used a stability-enhanced reward function to embed the flow instability (evaluated by an instability penalty) into the RL reward.
With the stability-enhanced reward and the \textit{wavemaker}-based probe placement, vortex shedding of confined cylinder wakes with different $\beta$ at higher $Re$ can be further suppressed with different degrees of success. 

In the end, we find that placing the velocity probes covering the \textit{wavemaker} is preferable in RL-based control of confined cylinder wakes to suppress the vortex shedding. As mentioned earlier, the \textit{wavemaker} region is a region where the flow is most sensitive to variation.
Our results show that placing the probes to cover the \textit{wavemaker} region yields a better performance of the RL than placing them otherwise. This can be interpreted as that the flow information (as the environment and state components in the RL framework) is more accurately detected by the probes if they are placed in this manner, so that the action can be more efficient in controlling the flow. Besides, the robustness has also been confirmed: as long as the main part of the \textit{wavemaker} is covered, small differences in the placement of probes do not affect the general good performance of RL. When the probes are properly placed, we also find that the RL-control policy can outperform a gradient-based optimisation method (optimised in a certain time-horizon) in the long run. When insufficient probes are used, the performance of the probe distributions considered in this work is scattered. Our heuristic approach is able to identify a good distribution as an initial strategy of the placement of the probes, but more systematic approaches should also be adopted, such as the one in \cite{paris2021robust} in order to obtain more desirable results. Combining this heuristic result with the optimisation method thus may be promising.

In this study, the best policy trained in a fixed episode number (500) is used, which has shown a good convergence since there is only one independent actuator.
For policy-gradient RL methods such as PPO, the policy network integrates over both state and action spaces, and increasing the number of actuators/actions may require much more training episodes.
Thus, the exploitation performance may be significantly decreased within a given learning budget, especially for cases using computationally expensive three-dimensional simulations.
In this circumstance, the deterministic policy gradient method~\citep{silver2014deterministic} might be more efficient since the policy merely integrates over the state space.
Besides, proper utilising the knowledge of flow physics in a control agent can simplify its structure without losing performance (see the phase control in the drag reduction of a bluff body in~\citet{Pastoor2008}, which is mostly based on the understanding of decoupling shear layer development and wake processes).
More research on the coupling of fluid mechanics and RL-based control is needed to improve efficiency, effectiveness, and robustness in real-world complex applications.\\

\noindent Declaration of Interests. The authors report no conflict of interest.

\begin{acknowledgments}
The work is supported by a Tier 2 grant from the Ministry of Education, Singapore (R-265-000-661- 112). We acknowledge the computational resources provided by the National Supercomputing Centre of Singapore.
\end{acknowledgments}

\begin{appendix}

\section{Open-source code}\label{AppendixA}

The code for the RL-based flow control in this study is open-source as a GitHub repository: \url{https://github.com/npuljc/RL_control_Nek5000}.
It is developed by referring to the open-source RL-based control code of~\citet{Rabault2019}, so they have the same structure and both are based on the PPO agent implemented in \texttt{Tensorforce}.
The main difference is the simulation environment, which is based on a SEM solver Nek5000 (Version 19.0, \citet{nek5000}) in our repository.
Another difference is that the reward function in our code is defined by the shedding energy (Eq.~\ref{eq_shedding_energy}), and there is an option to use the stability-enhanced reward (Eq.~\ref{eq_reward_growthrate}).
More details can be found in the \texttt{Readme} file of the repository.

\section{Dynamic mode decomposition}\label{AppendixB}

In our analysis, we have also utilised the dynamic mode decomposition (DMD) in order to further probe the flow instability when the conventional linear stability analysis is hard to apply. When the flow rate of the synthetic jet is zero (i.e, the control is off), the conventional linear stability analysis can be applied properly. However, when the control is on and the synthetic jet flow rates are varying, it turns out to be difficult to apply the conventional linear stability analysis and in this case we use DMD.

As first proposed by~\cite{SCHMID2010,Rowley2009}, DMD is a data-based method in the analysis of the time evolution of fluid flows. We briefly summarise the algorithm and refer the reader to \cite{SCHMID2010} for the complete theory.
For a series of flow snapshots $\boldsymbol{V}_1^{N_\text{snapshot}} = \boldsymbol{v}_1, \boldsymbol{v}_2, \dots, \boldsymbol{v}_{N_\text{snapshot}}$ that is generated with a time interval of $\Delta t$,
one assumes that a linear mapping $\boldsymbol{A}$ connects the flow field $\boldsymbol{v}_i$ to the subsequent flow field $\boldsymbol{v}_{i+1}$, that is, $\boldsymbol{v}_{i+1} = \boldsymbol{A} \boldsymbol{v}_i$.
Then, we have $\boldsymbol{A}\boldsymbol{V}_1^{{N_\text{snapshot}}-1} = \boldsymbol{V}_1^{{N_\text{snapshot}}-1}\boldsymbol{S} + \boldsymbol{r}\boldsymbol{e}_{{N_\text{snapshot}}-1}^{T}$, where $\boldsymbol{r}$ is the residual vector and $\boldsymbol{e}_{{N_\text{snapshot}}-1} \in \mathbb{R}^{{N_\text{snapshot}}-1}$ is a $({N_\text{snapshot}}-1)\text{th}$ unit vector.
The eigenvalues of $\boldsymbol{S}$ then approximate some of the eigenvalues of $\boldsymbol{A}$.
To improve the robustness, we use an implementation based on eigenvalue decomposition of a `full' matrix $\hat{\boldsymbol{S}}$, which is related to $\boldsymbol{S}$ via a similarity transformation.
\begin{equation}
  \hat{\boldsymbol{S}} = \boldsymbol{U} \boldsymbol{V}_2^{N_\text{snapshot}} \boldsymbol{W} \boldsymbol{\Sigma}^{-1},
\end{equation}
where $\boldsymbol{U}$, $\boldsymbol{W}$, and $\boldsymbol{\Sigma}$ are obtained by performing a singular value decomposition of $\boldsymbol{V}_1^{{N_\text{snapshot}}-1} = \boldsymbol{U} \boldsymbol{\Sigma} \boldsymbol{W}^{H}$.
The dynamic modes are $\boldsymbol{\Phi}_i = \boldsymbol{U}  \boldsymbol{y}_i$, where $\boldsymbol{y}_i$ is the $i\text{th}$ eigenvector of $\hat{\boldsymbol{S}}$, i.e.,
\begin{equation}
  \hat{\boldsymbol{S}} \boldsymbol{y}_i = \mu_i \boldsymbol{y}_i .
\end{equation}
The growth rates and frequencies of the modes can be obtained by the logarithmic mapping of corresponding DMD eigenvalues.
For the $i\text{th}$ mode with eigenvalue $\mu_i$, the the frequency $f_i = \text{Im}({\ln{\mu_i}})/2\pi \Delta t$ and growth rate $g_i =\text{Re}({\ln{\mu_i}})/\Delta t$.

\section{Gradient-based optimisation of the control policy for synthetic jet flow rates}\label{AppendixC}

For reference, gradient-based optimisation is used to solve the optimal control policy in a given control horizon of 32 time units, which equals to two episode in RL-based control.
The sequential least-squares programming (SLSQP) algorithm implemented in pyOptSparse~\citep{Perez2012a,Wu2020}(https://github.com/mdolab/pyoptsparse) is used to minimize the total shedding energy in the control horizon.
The shedding energy is evaluated using Eq.~\ref{eq_shedding_energy} with a reference of the SFD base flow.
The finite difference method implemented in pyOptSparse is used to solve the gradient of the shedding energy with respect to each control variable.
As mentioned above, the synthetic jet flow rate can be adjusted every 0.2 time unit, and this means that we have 160 independent control variables in the optimisation.
To reduce the computational cost in gradient evaluations, the 160 jet flow rates are parameterized by a cubic spline with 80 evenly distributed knots.
Referring the RL-based control policies, a cubic spline with 80 knots can provide enough degrees of freedom for the control of synthetic jet flow rates in this problem.
Thus, the gradient-based optimisation is actually subject to 80 design variables.
We summarize this optimisation problem in Table~\ref{tbopt}.
\begin{table}
  \centering
  \caption{optimisation problem statement of the synthetic jet flow rate control policy}
  \begin{tabular}{l*{6}{c}r}
  \hline
                   & Functions &  Quantity & Description \\
  \hline
  minimize         & $\sum_{t=0}^{t=32} e_s^{(t)}$       &   1     &  Total shedding energy in a control horizon of 32 time units\\
  $w.r.t.$  & $k_i$    &   80     &  Cubic spline knots for jet flow rates \\
  \hline
  \end{tabular}
  \label{tbopt}
\end{table}

\section{RL control of cylinder wakes with different $\beta$}\label{AppendixD}

\begin{figure}
  \centering
  \includegraphics[width=\textwidth]{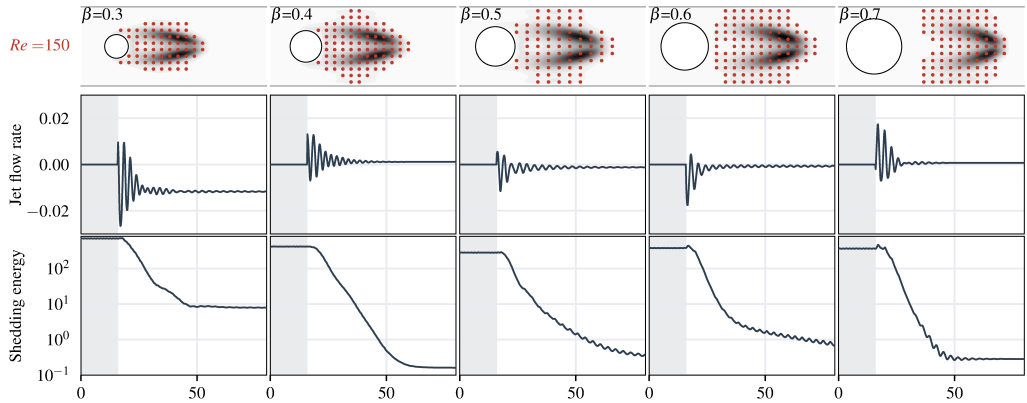}
  \caption{RL control performance of cylinder wakes with different $\beta$ at $Re=150$. The probes are placed based on the \textit{wavemaker}-based criterion.}
  \label{fg_diffprobesMoreRe150}
\end{figure}

To see if the \textit{wavemaker}-based probe placement criterion works with the change of the blockage ratio, we investigate the influence of different blockage ratios on RL-based control performance at $Re = 150$, and probes for RL control are adjusted based on the corresponding \textit{wavemaker} region of the confined wake. 
The results are shown in figure~\ref{fg_diffprobesMoreRe150}.

\end{appendix}

\bibliographystyle{jfm}
\bibliography{main}

\end{document}